\documentclass[twocolumn]{aastex63}
\accepted{2022 February 8th}

\usepackage{natbib,aas_macros,amsmath}
\usepackage{multirow,color}
\usepackage{natbib}

\newcommand{\carcsec}{$\!\!\arcsec$}
\newcommand{\m}[1]{\mathrm{#1}}

\newcommand{\redc}[1]{\textcolor{black}{#1}}
\newcommand{\redcc}[1]{\textcolor{black}{#1}}

\usepackage{booktabs}

\begin{document}
\shortauthors{Harikane et al.}

\shorttitle{
$z>12$ Dropouts
}

\title{
A Search for $H$-Dropout Lyman Break Galaxies at \redc{$z\sim12-16$}
}

\email{hari@icrr.u-tokyo.ac.jp}
\author[0000-0002-6047-430X]{Yuichi Harikane}
\affiliation{Institute for Cosmic Ray Research, The University of Tokyo, 5-1-5 Kashiwanoha, Kashiwa, Chiba 277-8582, Japan}
\affiliation{Department of Physics and Astronomy, University College London, Gower Street, London WC1E 6BT, UK}

\author[0000-0002-7779-8677]{Akio K. Inoue}
\affiliation{Department of Physics, School of Advanced Science and Engineering, Faculty of Science and Engineering, Waseda University, 3-4-1 Okubo, Shinjuku, Tokyo 169-8555, Japan}
\affiliation{Waseda Research Institute for Science and Engineering, Faculty of Science and Engineering, Waseda University, 3-4-1 Okubo, Shinjuku, Tokyo 169-8555, Japan}

\author[0000-0003-4985-0201]{Ken Mawatari}
\affiliation{National Astronomical Observatory of Japan, 2-21-1 Osawa, Mitaka, Tokyo 181-8588, Japan}

\author[0000-0002-0898-4038]{Takuya Hashimoto}
\affiliation{Tomonaga Center for the History of the Universe (TCHoU), Faculty of Pure and Applied Science, University of Tsukuba, Ibaraki, 305-8571, Japan}

\author[0000-0002-7738-5290]{Satoshi Yamanaka}
\affiliation{General Education Department, National Institute of Technology, Toba College, 1-1, Ikegami-cho, Toba, Mie 517-8501, Japan}
\affiliation{Research Center for Space and Cosmic Evolution, Ehime University, 2-5, Bunkyo-cho, Matsuyama, Ehime 790-8577, Japan}
\affiliation{Waseda Research Institute for Science and Engineering, Faculty of Science and Engineering, Waseda University, 3-4-1 Okubo, Shinjuku, Tokyo 169-8555, Japan}

\author[0000-0001-7440-8832]{Yoshinobu Fudamoto}
\affiliation{Waseda Research Institute for Science and Engineering, Faculty of Science and Engineering, Waseda University, 3-4-1 Okubo, Shinjuku, Tokyo 169-8555, Japan}
\affiliation{National Astronomical Observatory of Japan, 2-21-1 Osawa, Mitaka, Tokyo 181-8588, Japan}

\author[0000-0002-4559-6157]{Hiroshi Matsuo}
\affiliation{National Astronomical Observatory of Japan, 2-21-1 Osawa, Mitaka, Tokyo 181-8588, Japan}
\affiliation{Department of Astronomical Science, The Graduate University for Advanced Studies (SOKENDAI), 2-21-1 Osawa, Mitaka, Tokyo 181-8588, Japan}

\author[0000-0003-4807-8117]{Yoichi Tamura}
\affiliation{Division of Particle and Astrophysical Science, Graduate School of Science, Nagoya University, Nagoya 464-8602, Japan}

\author[0000-0001-8460-1564]{Pratika Dayal}
\affiliation{Kapteyn Astronomical Institute, University of Groningen, P.O. Box 800, 9700 AV Groningen, the Netherlands}

\author[0000-0003-3466-035X]{L. Y. Aaron Yung}
\affiliation{Astrophysics Science Division, NASA Goddard Space Flight Center, Greenbelt, MD 20771, USA}

\author[0000-0003-3760-461X]{Anne Hutter}
\affiliation{Kapteyn Astronomical Institute, University of Groningen, P.O. Box 800, 9700 AV Groningen, the Netherlands}

\author[0000-0001-9879-7780]{Fabio Pacucci}
\affil{Center for Astrophysics $\vert$ Harvard \& Smithsonian, Cambridge, MA 02138, USA}
\affil{Black Hole Initiative, Harvard University, Cambridge, MA 02138, USA}

\author[0000-0001-6958-7856]{Yuma Sugahara}
\affil{Waseda Research Institute for Science and Engineering, Faculty of Science and Engineering, Waseda University, 3-4-1 Okubo, Shinjuku, Tokyo 169-8555, Japan}
\affil{National Astronomical Observatory of Japan, 2-21-1 Osawa, Mitaka, Tokyo 181-8588, Japan}

\author[0000-0002-6610-2048]{Anton M. Koekemoer}
\affiliation{Space Telescope Science Institute, 3700 San Martin Dr., Baltimore, MD 21218, USA}

\begin{abstract}
We present two bright galaxy candidates at \redc{$z\sim12-13$} identified in our $H$-dropout Lyman break selection with $2.3\ \m{deg^2}$ near-infrared deep imaging data.
These galaxy candidates, selected after careful screening of foreground interlopers, have spectral energy distributions showing a sharp discontinuity around 1.7 $\mu$m, a flat continuum at $2-5$ $\mu$m, and non-detections at $<1.2$ $\mu$m in the available photometric datasets, all of which are consistent with $z>12$ galaxy.
An ALMA program targeting one of the candidates shows a tentative $4\sigma$ {\sc [Oiii]}88$\mu$m line at $z=13.27$, in agreement with its photometric redshift estimate.
The number density of the \redc{$z\sim12-13$} candidates is comparable to that of bright $z\sim10$ galaxies, and is consistent with a recently proposed double power-law luminosity function rather than the Schechter function, indicating little evolution in the abundance of bright galaxies from $z\sim4$ to $13$.
Comparisons with theoretical models show that the models cannot reproduce the bright end of rest-frame ultraviolet luminosity functions at $z\sim10-13$.
Combined with recent studies reporting similarly bright galaxies at $z\sim9-11$ and mature stellar populations at $z\sim6-9$, our results indicate the existence of a number of star-forming galaxies at $z>10$, which will be detected with upcoming space missions such as {\it James Webb Space Telescope}, {\it Nancy Grace Roman Space Telescope}, and {\it GREX-PLUS}.
\end{abstract}


\section{Introduction}\label{ss_intro}

Observing the first galaxy formation is one of the main goals in the modern astronomy.
One of the most straightforward approaches to achieve this goal is to observe forming galaxies directly in the early universe.
Large telescopes currently in operation have yielded the most distant objects so far.
These highest redshift objects have posed various interesting questions for astronomy.
For example, the most distant quasars at $z>7$ raised a serious problem to form blackholes as massive as $\sim10^9\ M_\odot$ in the limited cosmic time \citep[e.g.,][]{2011Natur.474..616M,2018Natur.553..473B,2020ApJ...897L..14Y,2021ApJ...907L...1W}.
Thus, searching for the most distant objects is not only the simplest frontier of the knowledge of human beings but also has a great power to reveal the formation physics of various objects in the early universe \citep[e.g., see review by][]{2016ARA&A..54..761S,2018PhR...780....1D,2021arXiv211013160R}.

The current record of the highest redshift galaxy spectroscopically confirmed is GN-z11 at $z\sim11$ measured with detections of the Lyman break and rest-frame ultraviolet (UV) metal lines \citep{2016ApJ...819..129O,2021NatAs...5..256J}.
A major surprise of GN-z11 is its remarkably high luminosity, $M_\m{UV}=-22.1$ mag.
Given that it is not gravitationally-lensed, GN-z11 is located in the brightest part of the rest-frame UV luminosity function.
Although the narrow field-of-view (FoV) of {\it Hubble Space Telescope (HST)}/Wide Field Camera 3 (WFC3) in the near-infrared has limited the imaging survey areas to $<1\ \m{deg^2}$, several studies using {\it HST} report very luminous Lyman break galaxies (LBGs) at $z\sim9-10$ more frequently than the expectation from a Schechter-shape luminosity function (e.g., \citealt{2018ApJ...867..150M,2021arXiv210613813F}, see also \citealt{2021arXiv210606544R}).
More statistically robust results have come from a few square-degree near-infrared imaging surveys with Visible and Infrared Survey Telescope for Astronomy (VISTA) and UK Infrared Telescope (UKIRT) such as UltraVISTA \citep{2012A&A...544A.156M}, the UKIRT InfraRed Deep Sky Surveys (UKIDSS, \citealt{2007MNRAS.379.1599L}), and the VISTA Deep Extragalactic Observation (VIDEO) Survey \citep{2013MNRAS.428.1281J}.
These surveys have revealed that the UV luminosity functions at $z\sim9-10$ are more consistent with a double power-law than a standard Schechter function \citep{2017ApJ...851...43S,2019ApJ...883...99S,2020MNRAS.493.2059B}.
Previous studies also report similar number density excesses beyond the Schechter function at $z\sim4-7$ \citep{2018PASJ...70S..10O,2018ApJ...863...63S,2020MNRAS.494.1771A,2021arXiv210801090H}, implying little evolution of the number density of bright galaxies at $z\sim4-10$ \citep{2020MNRAS.493.2059B,2021arXiv210801090H}.
Although spectroscopic observations are required to confirm these results, the studies indicate that there are a larger number of luminous galaxies at $z\sim9-11$ than previously thought, which formed in the early universe of $z>10$.

In addition to these observations of bright galaxies at $z\sim9-11$, several studies independently suggest the presence of star-forming galaxies in the early universe even at $z\sim15$.
A candidate for a $z\sim12$ galaxy is photometrically identified in very deep {\it HST}/WFC3 images obtained in the Hubble Ultra Deep Field 2012 (UDF12) campaign \citep{2013ApJ...763L...7E}.
Balmer breaks identified in $z=9-10$ galaxies indicate mature stellar populations whose age is $\sim300-500$ Myr, implying early star formation at $z\sim14-15$ (\citealt{2018Natur.557..392H}, \citealt{2021MNRAS.505.3336L}, see also \citealt{2020MNRAS.497.3440R}).
An analysis of passive galaxy candidates at $z\sim6$ reports that their stellar population is dominated by old stars with ages of $\gtrsim700$ Myr, consistent with star formation activity at $z>14$ \citep{2020ApJ...889..137M}.

Motivated by these recent works, we search for $H$-band dropout ($H$-dropout) LBGs whose plausible redshifts are \redc{$z\sim12-16$}.\footnote{$H$-dropouts sources searched in this work are different from previously-studied dusty ``$H$-dropouts" at $z\sim3-6$ \citep[e.g.,][]{2019Natur.572..211W}.}
Given the observed number density of luminous galaxies at $z\sim9-10$ and its little redshift evolution from $z\sim4$ to $10$, it is possible that one to several \redc{$z\sim12-16$} galaxies will be found in currently available datasets obtained by surveys with large ground- and space-based telescopes.
This search for \redc{$z\sim12-16$} galaxies is important not only for understanding early galaxy formation, but also for designing survey strategies with upcoming space missions that will study the $z>10$ universe such as {\it James Webb Space Telescope (JWST)}.

\begin{deluxetable*}{ccccccccccccccccc}
\renewcommand{\arraystretch}{0.9}
\tablecaption{$5\sigma$ Limiting Magnitude of Imaging Data Used in This Study}
\tablehead{
\colhead{} & \colhead{} & \colhead{} & \colhead{} & \multicolumn{5}{c}{Subaru} & & \multicolumn{3}{c}{VISTA/UKIRT} & & \multicolumn{2}{c}{\it Spitzer}\\
\cline{5-9} \cline{11-13} \cline{15-16} 
\colhead{Field} & \colhead{\redc{R.A.}} & \colhead{\redc{decl.}} & \colhead{\redc{$A_\m{Survey}$}} & \colhead{$g$} & \colhead{$r$} & \colhead{$i$} & \colhead{$z$} & \colhead{$y$} & & \colhead{$J$}& \colhead{$H$} & \colhead{$K_\m{s}(K)$} & & \colhead{$[3.6]$} & \colhead{$[4.5]$}
}
\startdata
UD-COSMOS & \redcc{10:00:10} & \redcc{$+$02:12:41} & \redc{1.5 $\m{deg}^2$} & 26.9 & 26.6 & 26.8 & 26.6 & 25.9 & & 25.6/24.5 & 25.2/24.1 & 24.9/24.5 & & 25.1 & 24.9 \\
UD-SXDS & \redcc{02:17:48} & \redcc{$-$05:05:44} & \redc{0.8 $\m{deg}^2$} & 27.2 & 26.7 & 26.6 & 26.1 & 25.3 & & 25.6 & 25.1 & 25.3 & & 25.3 & 24.9
\enddata
\tablecomments{\redc{R.A. and decl. are the central coordinates of the survey field.} The $5\sigma$ limiting magnitudes are measured in $1.\carcsec5$, $2.\carcsec0$, and $3.\carcsec0$-diameter apertures in $grizy$, $JHK_s(K)$, and $[3.6][4.5]$ images, respectively, taken from \citet{2021arXiv210801090H}, release notes of UltraVISTA DR4\footnote{http://ultravista.org/release4/} (McCracken et al. 2012) and UKIDSS DR11\footnote{https://www.nottingham.ac.uk/astronomy/UDS/data/dr11.html} \citep{2007MNRAS.379.1599L}, and \citet{2018ApJ...859...84H}.
The values for the $JHK_\m{s}$ bands in the UD-COSMOS field represent limiting magnitudes in the ultra-deep and deep stripes.}
\label{tab_data}
\end{deluxetable*}

This paper is organized as follows.
We describe photometric datasets and a selection of \redc{$z\sim12-16$} galaxies in Section \ref{ss_data}, and ALMA follow-up observations for one of our candidates in Section \ref{ss_ALMA}.
Results of spectral energy distribution (SED) fitting and the UV luminosity function are presented in Sections \ref{ss_SEDfit} and \ref{ss_UVLF_SFRD}, respectively.
We discuss future prospects with space missions based on our results in Section \ref{ss_future}, and summarize our findings in Section \ref{ss_summary}.
Throughout this paper, we use the Planck cosmological parameter sets of the TT, TE, EE+lowP+lensing+ext result \citep{2016A&A...594A..13P}:
$\Omega_\m{m}=0.3089$, $\Omega_\Lambda=0.6911$, $\Omega_\m{b}=0.049$, $h=0.6774$, and $\sigma_8=0.8159$.
All magnitudes are in the AB system \citep{1983ApJ...266..713O}.

\section{Photometric Dataset and Sample Selection}\label{ss_data}

\subsection{Dataset}

We use deep and wide photometric datasets available in the COSMOS \citep{2007ApJS..172....1S} and SXDS \citep{2008ApJS..176....1F} fields.
The total survey area is about 2.3 deg$^2$, which is almost limited by the coverage of the deep near-infrared data.
Specifically, we use optical $grizy$ images obtained in the Hyper Suprime-Cam Subaru Strategic Program (HSC-SSP) Survey \citep{2018PASJ...70S...4A,2019PASJ...71..114A} public data release 2 (PDR2), near-infrared $JHK_\m{s}/K$ images of the UltraVISTA DR4 \citep{2012A&A...544A.156M} and UKIDSS UDS DR11 \citep{2007MNRAS.379.1599L} in the COSMOS and SXDS fields, respectively, and {\it Spitzer}/IRAC [3.6] and [4.5] images obtained in the {\it Spitzer} Large Area Survey with Hyper-Suprime-Cam (SPLASH).
Typical $5\sigma$ limiting magnitudes of these imaging data are presented in Table \ref{tab_data}.
Since the COSMOS field has the ultra-deep and deep stripes with different depths in the near-infrared images, we use the limiting magnitude of each stripe depending on the location of the source of interest.

\subsection{Selection of $H$-dropout Galaxies}\label{ss_select}

We construct multi-band photometric catalogs in the COSMOS and SXDS fields.
We start from $K_\m{s}(K)$-band detection catalogs made by the UltraVISTA (UKIDSS) team using {SExtractor} \citep{1996A&AS..117..393B}.
We select sources detected in the $K_\m{s}(K)$ bands at $>5\sigma$ levels and not detected in the $J$-band at $>2\sigma$ levels in a $2\arcsec$-diameter circular aperture.
Then we measure magnitudes of these sources in the $grizyJHK_\m{s}(K)$ images using a $2\arcsec$-diameter circular aperture centered at their coordinates in the catalogs.
Since point-spread functions (PSFs) of the {\it Spitzer}/IRAC [3.6] and [4.5] images are relatively large ($\sim1.\carcsec7$), source confusion and blending are significant for some sources.
To remove the effects of the neighbor sources on the photometry, we first generate residual IRAC images where only the sources under analysis are left by using {T-PHOT} \citep{2016A&A...595A..97M}, in the same manner as \citet{2018ApJ...859...84H,2019ApJ...883..142H}.
As high-resolution prior images in the {T-PHOT} run, we use HSC $grizy$ stacked images whose PSF is $\sim0.\carcsec7$.
Then we measure magnitudes in the IRAC images by using a $3\arcsec$-diameter apertures in the same manner as \citet{2018ApJ...859...84H}.
To account for the flux falling outside the aperture, we apply aperture corrections derived from samples of isolated point souces in \citet{2018ApJ...859...84H}.

\begin{figure*}
\centering
\begin{minipage}{0.49\hsize}
\begin{center}
\includegraphics[width=0.9\hsize, bb=6 11 358 358]{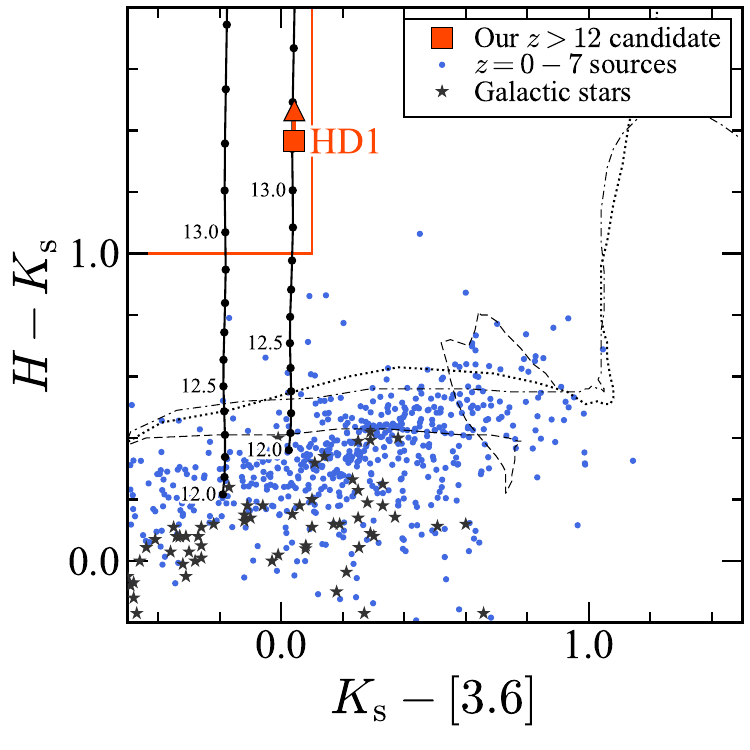}
\end{center}
\end{minipage}
\begin{minipage}{0.49\hsize}
\begin{center}
\includegraphics[width=0.9\hsize, bb=6 11 358 358]{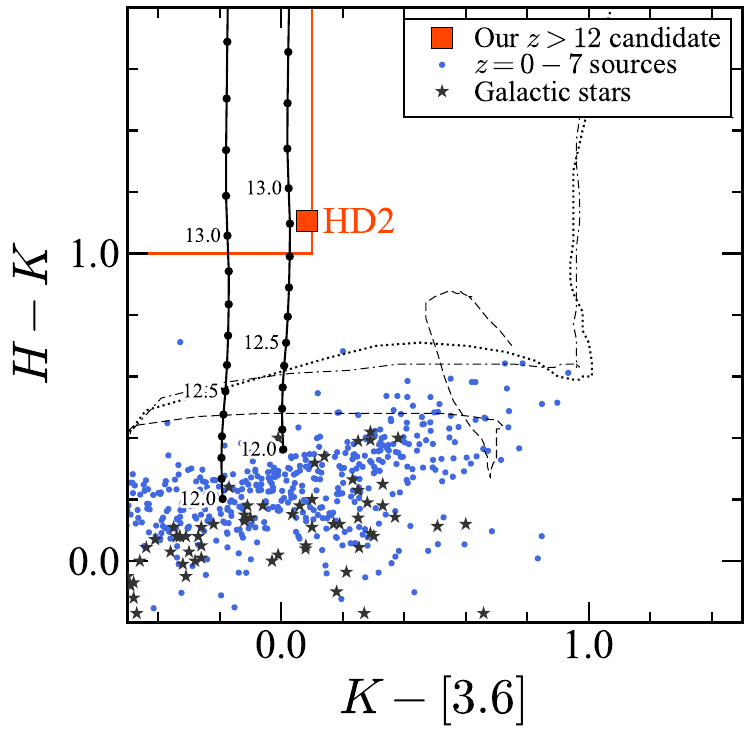}
\end{center}
\end{minipage}
\caption{Two-color diagrams to select $H$-dropout galaxies.
The left and right panels show two-color diagrams in the COSMOS and SXDS fields, respectively.
The red lines indicate color criteria that we use to select $H$-dropout galaxies (Equations (\ref{eq_color_1})-(\ref{eq_color_4})), and the red squares are the selected candidates, HD1 and HD2.
The black solid lines are colors of star-forming galaxies at $z\geq12$ calculated with {BEAGLE} \citep{2016MNRAS.462.1415C} with $\tau_\m{V}=0.0$ and $0.4$ (corresponding to UV spectral slopes of $\beta_\m{UV}\simeq-2.4$ and $-1.9$, respectively) as a function of redshift.
The circles on the line show their redshifts with an interval of $\Delta z=0.1$.
The blue circles are $z=0-7$ sources spectroscopically identified \citep{2016ApJS..224...24L,2018ApJS..235...36M}.
The dotted, dashed, and dot-dashed lines are, respectively, typical spectra of elliptical, Sbc, and irregular galaxies \citep{1980ApJS...43..393C} redshifted from $z=0$ to $z=7$.
The black stars indicate Galactic dwarf stars taken from \citet{2006ApJ...651..502P} and \citet{2011ApJS..197...19K}.
}
\label{fig_2color}
\end{figure*}

\begin{deluxetable*}{ccccccccccccccc}
\tablecaption{Photometry of Our $H$-Dropout Galaxy Candidates}
\tablehead{\colhead{Name} & \colhead{R.A.} & \colhead{Decl.} & \multicolumn{5}{c}{Subaru} & & \multicolumn{3}{c}{VISTA/UKIRT} & & \multicolumn{2}{c}{\it Spitzer}\\
\cmidrule{4-8} \cmidrule{10-12} \cmidrule{14-15}
\colhead{} & \colhead{} & \colhead{} & \colhead{$g$} & \colhead{$r$} & \colhead{$i$} & \colhead{$z$} & \colhead{$y$} & & \colhead{$J$}  & \colhead{$H$}  & \colhead{$K_\m{s}/K$} & & \colhead{$[3.6]$}  & \colhead{$[4.5]$} \\
\colhead{(1)}& \colhead{(2)}& \colhead{(3)}& \colhead{(4)} &  \colhead{(5)}& \colhead{(6)}& \colhead{(7)}& \colhead{(8)}& & \colhead{(9)}& \colhead{(10)}& \colhead{(11)}& & \colhead{(12)}& \colhead{(13)} }
\startdata
HD1 & 10:01:51.31 & 02:32:50.0 & $<27$ & $<32$ & $<46$ & $<63$ & $<157$ & & $<107$ & $<145$ & $510\pm93$ & &  $531\pm108$ & $494\pm136$\\
HD2 & 02:18:52.44 & -05:08:36.1 & $<25$ & $<33$ & $<46$ & $<68$ & $<133$ & & $<95$ & $296\pm76$ & $821\pm63$ & & $888\pm88$ & $1252\pm132$
\enddata
\tablecomments{(1) Name.
(2) Right ascension.
(3) Declination.
(4)-(13) Flux densities in nJy or $2\sigma$ upper limits.
}
\label{tab_photo}
\end{deluxetable*}

\begin{figure*}
\centering
\includegraphics[width=0.91\hsize, bb=7 7 864 286]{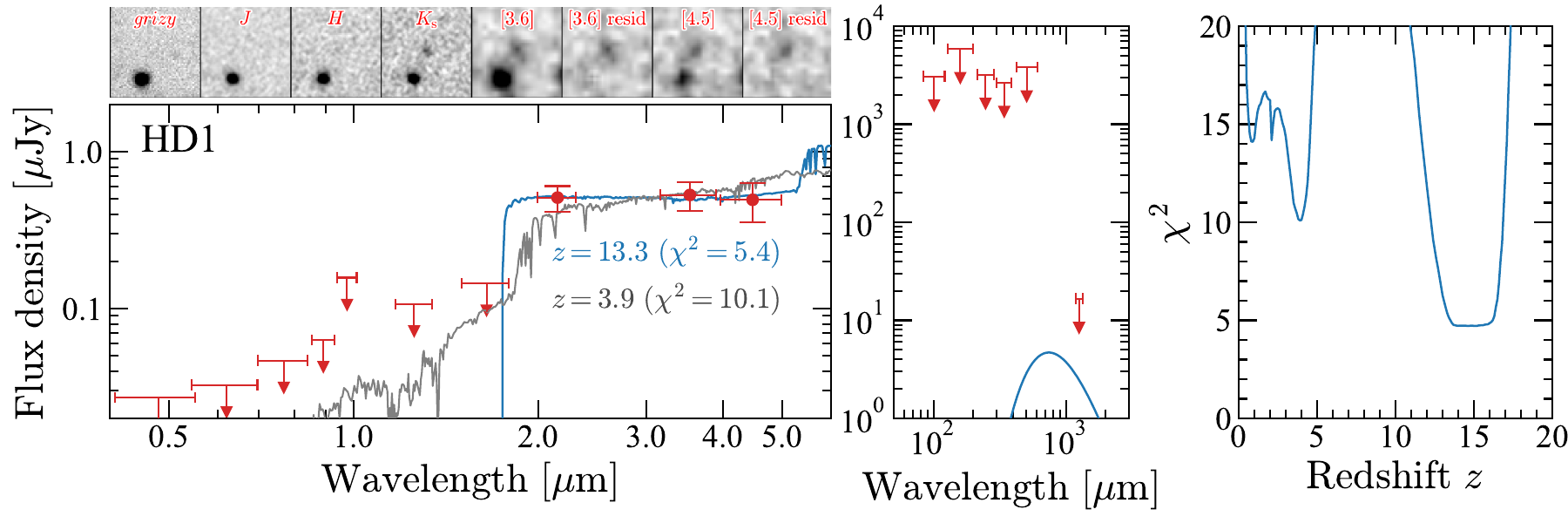}
\includegraphics[width=0.91\hsize, bb=7 7 864 290]{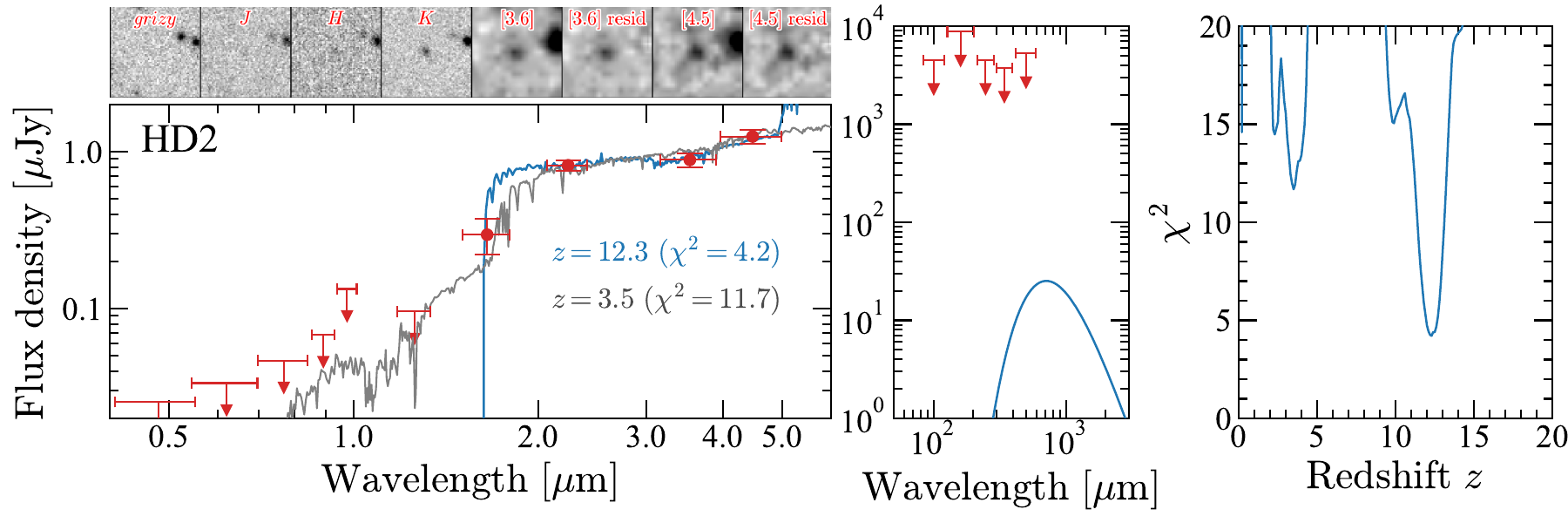}
\caption{
{\it Left:}
Optical-to-near-infrared SEDs of our $H$-dropout galaxy candidates, HD1 (top) and HD2 (bottom).
The red symbols with error-bars are measured flux densities or the 2$\sigma$ upper limits.
The blue curve shows the best-fit model of an LBG at $z>12$ in the SED fitting, and the gray curve shows a passive galaxy solution at $z\sim4$ (see Section \ref{ss_SEDfit}).
The upper panels show $10\arcsec\times10\arcsec$ images.
The ``[3.6] resid" and ``[4.5] resid" are residual images after subtracting nearby objects with {T-PHOT} \citep{2016A&A...595A..97M}.
{\it Middle:}
The same as the left panel but for the far-infrared-to-submm range.
The curve shows the modified black-body function with the temperature of 50 K and the emissivity index of $\beta_\m{dust}= 2.0$.
{\it Right:}
$\chi^2$ value as a function of the redshift.
The best-fit models are found at $z > 12$.
}
\label{fig_SED}
\end{figure*}

\begin{figure}
\centering
\includegraphics[width=0.87\hsize, bb=8 5 416 204,clip]{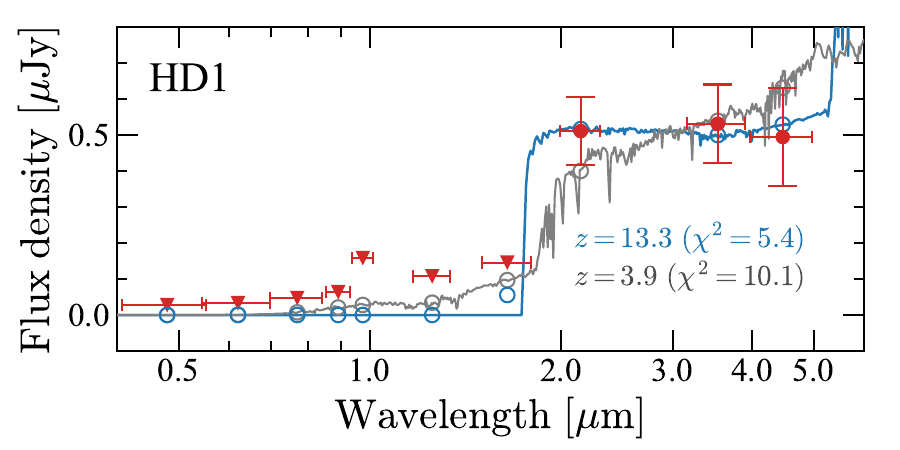}
\includegraphics[width=0.87\hsize, bb=8 5 416 209,clip]{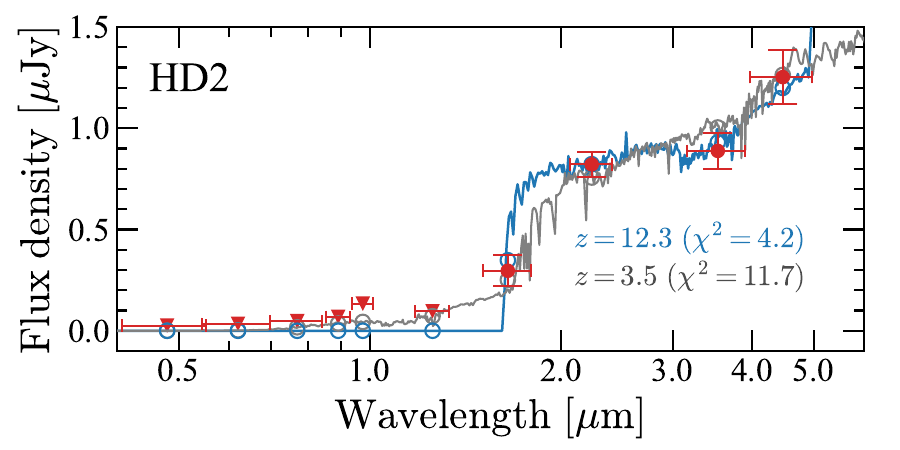}
\caption{
\redc{
Same as the left panels of Figure \ref{fig_SED} but with fluxes plotted in linear scale to compare the observed fluxes with the models.
The blue and grey open circles are fluxes of the models in each band.
}
}
\label{fig_SED_linear}
\end{figure}

To search for \redc{$z\sim12-16$} galaxies whose Lyman breaks are redshifted to $\sim1.6-2.1\ \mu\mathrm{m}$, we select $H$-dropout LBGs from the multi-band photometric catalogs constructed above.
We adopt the following color selection criteria in the COSMOS and SXDS fields, respectively:
\\
\noindent
{COSMOS:}
\begin{eqnarray}
H-K_\m{s}&>&1.0,\label{eq_color_1}\\
K_\m{s}-[3.6]&<&0.1,\label{eq_color_2}
\end{eqnarray}
\noindent
{SXDS:}
\begin{eqnarray}
H-K&>&1.0,\label{eq_color_3}\\
K-[3.6]&<&0.1.\label{eq_color_4}
\end{eqnarray}
As shown in Figure \ref{fig_2color}, these color criteria can select sources at \redc{$z>12$} while avoiding color tracks of $z=0-7$ galaxies and stellar sources.
To remove foreground interlopers, we exclude sources with detections at $>2\sigma$ levels in the $grizyJ$ band images.
Note that we use the same values for the color criteria in the COSMOS and SXDS fields.
Since the filter response profiles are different in the VISTA $JHK_\m{s}$ filters for the COSMOS field and the UKIRT $JHK$ filters for the SXDS field, the selection functions will not be identical.
We will account for this difference by separately evaluating the selection functions in the COSMOS and SXDS fields in Section \ref{ss_completeness}.

To remove foreground interlopers further, we conduct a photometric redshift analysis using {BEAGLE} \citep{2016MNRAS.462.1415C}.
We adopt a constant star formation history with the \citet{2003PASP..115..763C} initial mass function (IMF), stellar ages of $10^6$, $10^7$, and $10^8\ \m{yr}$, metallicities of 0.2 and 1 $Z_\odot$, and the \citet{2000ApJ...533..682C} dust attenuation law with the $V$-band optical depth of $\tau_\m{V}=0-2$ (steps of 0.2) to account for very dusty low redshift interlopers.
We select objects whose high redshift solution is more likely than the low redshift ones at a $>2\sigma$ level, corresponding to $\Delta \chi^2>4.0$, in the same manner as \citet{2020MNRAS.493.2059B}.
Then we visually inspect images and SEDs of the selected sources to remove spurious sources, sources affected by bad residual features in the {T-PHOT}-made IRAC images, and extremely red sources (e.g., $K_\m{s}(K)-[4.5]\gtrsim1$) that are not likely to be \redc{$z\sim12-16$} galaxies.

After these careful screening processes, we finally identify two \redc{$z\sim12-16$} galaxy candidates, HD1 and HD2, \redc{in the COSMOS and SXDS fields, respectively}.
Figures \ref{fig_SED} \redc{and \ref{fig_SED_linear}} \redcc{show} images and SEDs of HD1 and HD2, and Table \ref{tab_photo} summarizes their measured fluxes.
HD1 and HD2 are spatially isolated from other nearby sources, ensuring the robustness of the photometry.

HD1 is also found in the COSMOS2020 catalog \citep{2021arXiv211013923W}.
However, the photometric redshift of HD1 is 3.6 in the COSMOS2020 catalog.
This is due to the difference in the measured magnitudes in the IRAC images.
Our measured magnitudes are 24.6 and 24.7 mag in the [3.6] and [4.5] images, respectively, while 24.2 and 23.9 mag (24.4 and 24.1 mag) are cataloged in the COSMOS2020 CLASSIC (FARMER) catalog with very small flux errors of $3-12\ \m{nJy}$.
If we re-measure magnitudes by using a larger aperture in the original IRAC images before the {T-PHOT} run, the magnitudes become brighter due to a neighboring source located $\sim3.\carcsec5$ from HD1.
We additionally test with deeper SMUVS images \citep{2018ApJS..237...39A}.
We carefully measure the magnitudes in the SMUVS images and still find that the best photometric redshift for HD1 is $z>12$.
Magnitudes in the other bands including the $K_\m{s}$-band  in the COSMOS2020 catalog are consistent with our measurements, although their flux errors are much smaller than ours.
Thus in this study, we adopt our measured magnitudes for HD1.

\begin{figure*}
\centering
\includegraphics[width=0.9\hsize, bb=10 13 708 278]{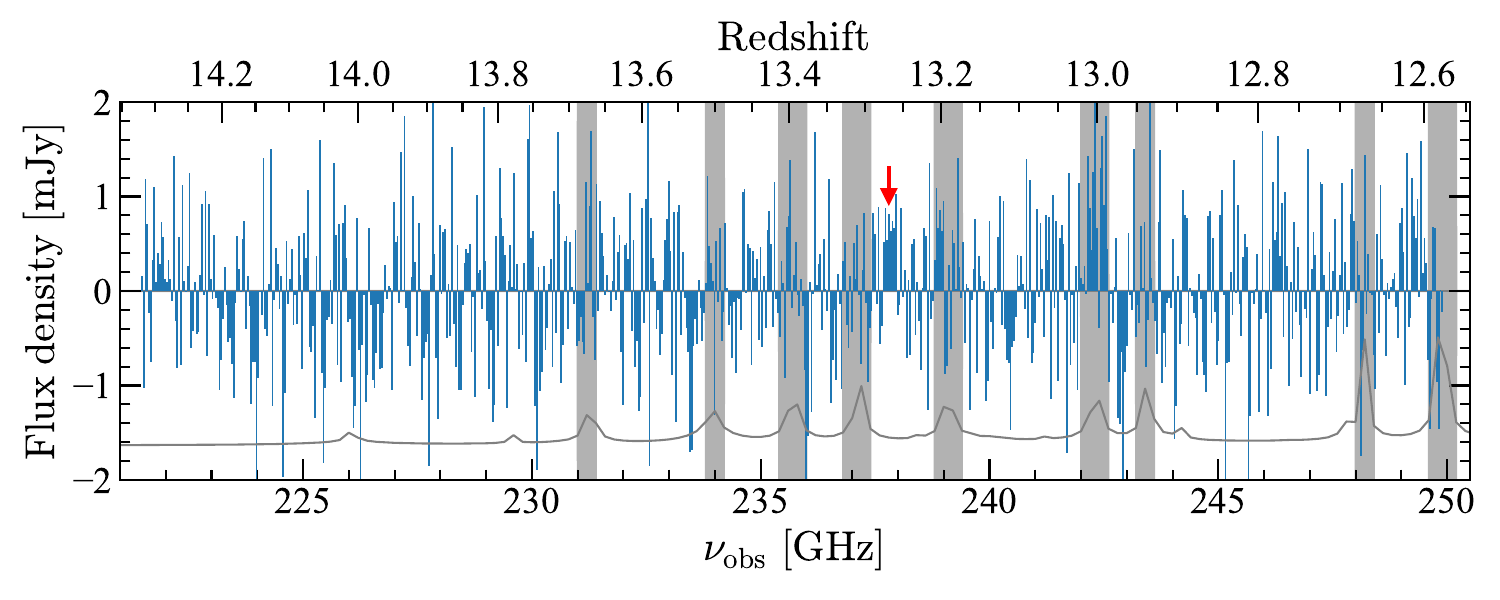}
\caption{Full ALMA spectrum of HD1.
This spectrum is extracted from a 1.\carcsec0-radius circular aperture centered on the coordinate of HD1.
No obvious emission line is identified at $>5\sigma$, but there is a $4\sigma$ line-like feature at 237.8 GHz (the red arrow), where no severe atmospheric O$_3$ absorption exists (the gray shades).
}
\label{fig_ALMA_full}
\end{figure*}

\redc{HD1 and HD2 will be observed in a {\it JWST} program \citep[GO-1740,][]{2021jwst.prop.1740H}.
In addition to these two candidates, the program will target another source, HD3 ($\m{R.A.}$=02:16:54.48, $\m{Decl.}$=-05:09:37.1), which is also a good candidate for a \redc{$z\sim12-16$} galaxy with its prominent break with $H-K>1.2$ and the best photometric redshift of $z_\m{phot}=14.6$.
However, due to its relatively red color with $K-[3.6]=0.2\pm0.2$, HD3 is not included in our final sample in this paper.}

\begin{figure}
\centering
\includegraphics[width=0.95\hsize, bb=7 0 354 278,clip]{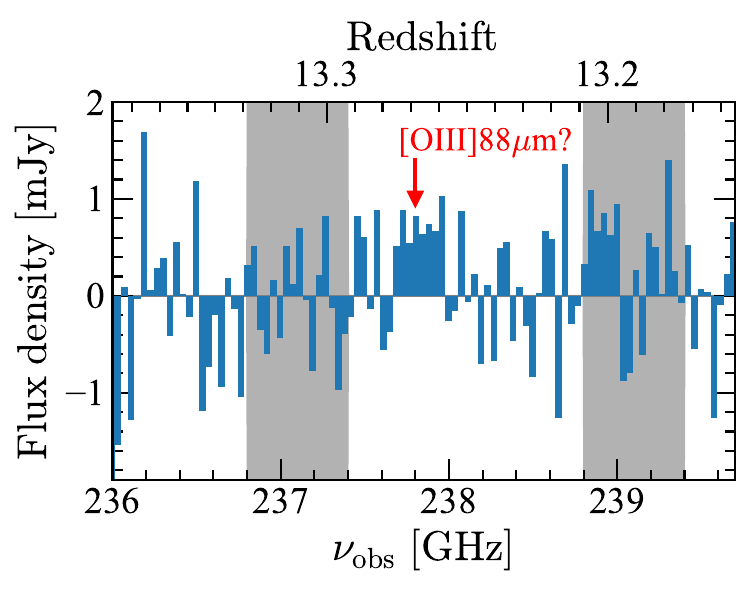}
\includegraphics[width=0.45\hsize, bb=6 0 148 148,clip]{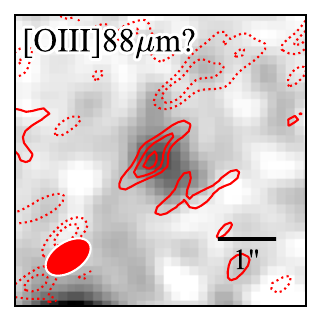}
\caption{
{\it Top:}
ALMA spectrum showing the $4\sigma$ line-like feature at 237.8 GHz.
This hints for the {\sc[Oiii]}88$\mu$m line at $z = 13.27$.
{\it Bottom:}
Integrated intensity of the $4\sigma$ feature in HD1 overlaid on the VISTA $K_\m{s}$ band image.
This moment 0 map is made with the CASA task {\tt immoments}, by integrating over 700 $\m{km\ s^{-1}}$ covering most of the velocity range of the line emission ($>1.5\times\m{FWHM}$).
The solid (dotted) lines show $+1.5$, $+2.5$, and $+3.5\sigma$ ($-1.5$, $-2.5$, and $-3.5\sigma$) contours.
The emission is co-spatial with the rest-frame UV emission in the $K_\m{s}$-band image.
}
\label{fig_ALMA}
\end{figure}

\section{ALMA Follow-up Observation}\label{ss_ALMA}

We observed one of the candidates, HD1, in an ALMA Director's Discretionary Time (DDT) program (2019.A.00015.S, PI: A. K. Inoue).
Following successful detections of {\sc[Oiii]}88$\mu$m emission lines in high redshift galaxies (e.g., \citealt{2016Sci...352.1559I}, \citealt{2017ApJ...837L..21L,2021MNRAS.505.3336L}, \citealt{2017A&A...605A..42C}, \citealt{2018Natur.553...51M}, \citealt{2018Natur.557..392H,2019PASJ...71...71H}, \citealt{2018ApJ...869L..22W}, \citealt{2019ApJ...874...27T}, \citealt{2020ApJ...896...93H}, see also \citealt{2014ApJ...780L..18I}), we conducted a spectral scan targeting the {\sc[Oiii]}88$\mu$m line using four tuning setups with the Band 6 covering the redshift range of $12.6<z<14.3$.
The antenna configurations were C43-2, C43-3, C43-4, and C43-5, and typical beam size is $\sim0.\carcsec4-0.\carcsec7$.
We used four spectral windows with 1.875 GHz bandwidths in the Frequency Division Mode and the total bandwidth of 7.5 GHz in one tuning setup.
The velocity resolution was set to $\sim10\ \m{km\ s^{-1}}$.
The data were reduced and calibrated using the Common Astronomy Software (CASA; \citealt{2007ASPC..376..127M}) pipeline version 5.4.0 in the general manner with scripts provided by the ALMA observatory.

Figure \ref{fig_ALMA_full} shows the obtained spectrum for HD1 extracted from a 1.\carcsec0-radius circular aperture.
Although there is no signal at a $>5\sigma$ level, we find a $4\sigma$ tentative line-like feature around 238 GHz.
As shown in the top panel of Figure \ref{fig_ALMA}, this feature is at 237.8 GHz, and the significance level of the peak intensity is $3.8\sigma$ in the moment 0 map shown in the bottom panel.
Although there are some other line-like features (e.g., 246.3 GHz), the feature at 237.8 GHz has the highest signal-to-noise ratio among the ones in the frequencies free from severe atmospheric O$_3$ absorption.
If this feature is the {\sc[Oiii]}88$\mu$m emission line, the redshift of HD1 is $z=13.27$, in good agreement with the photometric redshift estimate.
A relatively broad line width ($\sim400\ \m{km\ s^{-1}}$ in a full width at half maximum; FWHM) is in fact comparable to similarly bright LBGs at $z\sim6$ \citep{2020ApJ...896...93H}. 
The emission feature is co-spatial with the rest-frame UV emission in the $K_\m{s}$-band image (the bottom panel of Figure \ref{fig_ALMA}).
The integrated line flux is $0.24\pm0.06\ \m{Jy\ km\ s^{-1}}$ or $(1.9\pm0.5)\times10^{-18}\ \m{erg\ s^{-1}\ cm^{-2}}$, and the line luminosity is $L_\m{[OIII]}\simeq3.3\times10^8\ L_\odot$ if $z=13.27$ is assumed.

The line luminosity is very small compared to the UV luminosity.
Since the UV luminosity of HD1 is $L_{\rm UV}=4.8\times10^{11}$ $L_\odot$, the {\sc [Oiii]}-to-UV luminosity ratio is $L_\m{[OIII]}/L_\m{UV}\sim7\times10^{-4}$.
This ratio is the smallest among the galaxies observed in {\sc [Oiii]}88$\mu$m emissions in the reionization epoch as well as in the local Universe so far (e.g.,  \citealt{2016Sci...352.1559I,2021A&A...646A..26B}).
Since the {\sc [Oiii]}-to-UV ratio depends on the oxygen abundance \citep{2020ApJ...896...93H}, this low ratio indicates a metallicity as low as $\sim0.01-0.1\ Z_\odot$.

Another possibility is that the ALMA line scan just missed the true emission line and the redshift is out of the range of $12.6<z<14.3$.
As we will see in Section \ref{ss_completeness}, the redshift selection function is as broad as $12<z<17$.
For HD1, the lower redshift case ($z<12.6$) is not very favored by the SED fitting, but the higher redshift case ($z>14.3$) is still equally likely, as we discuss later in Section \ref{ss_SEDfit}.
Therefore, additional spectroscopic data are highly desired to confirm redshifts of HD1 and HD2.
We plan to conduct follow-up observations for the tentative signal in HD1 and to newly obtain spectroscopic data for HD2 in ALMA cycle 8 (2021.1.00207.S, PI: Y. Harikane).
We will also observe these candidates with {\it JWST} (GO-1740, \citealt{2021jwst.prop.1740H}), which allows us to examine a wider redshift range than ALMA.

The dust continuum of HD1 remains non-detection, which is consistent with the low metallicity interpretation from the low $L_\m{[OIII]}/L_\m{UV}$ ratio discussed above.
The obtained $1\sigma$ noise level is $8~\mu$Jy beam$^{-1}$.
Assuming that HD1 is not resolved in this observation, we obtain the $3\sigma$ upper limit on the dust continuum of $<24~\mu$Jy.

\section{SED Fitting}\label{ss_SEDfit}

To examine the photometric redshifts of HD1 and HD2 more carefully, we perform a comprehensive SED fitting analysis from optical to sub-millimeter (sub-mm) wavelength using {\tt PANHIT} \citep{2020IAUS..341..285M}.
{\tt PANHIT} takes the energy conservation of the dust absorption in the rest-frame UV to near-infrared range and the emission in the far-infrared to sub-mm range into account.
{\tt PANHIT} deals with the upper limits for non-detection bands, following the probability distribution function formula proposed by \cite{2012PASP..124.1208S}.
We adopt $1\sigma$ for the upper bound of the integral of the probability distribution.
In addition to the fluxes in $grizyJHK_\m{s}(K)[3.6][4.5]$ measured in Section \ref{ss_select}, we utilize far-infrared and sub-mm data of the {\it Herschel} survey \citep{2012MNRAS.424.1614O} of HD1 and HD2, and the ALMA data obtained for HD1 (see Section \ref{ss_ALMA}).
Since dust continua of HD1 and HD2 are not detected in these data, we use the upper limits for the SED fitting.

We assume a delayed-$\tau$ model for the star formation history (SFH) covering a wide range of histories including a short time-scale burst, rising, declining, and almost constant cases \citep{2014ApJS..214...15S}.
It is important to include passive galaxy models because the red $H-K_\m{s}(K)$ color can be produced by the Balmer break as well as the Lyman break. 
This may be a major contamination case in our $H$-dropout selection.
Template spectra include the BC03 stellar population synthesis model \citep{2003MNRAS.344.1000B} with the \citet{2003PASP..115..763C} IMF of 0.1--100 $M_\odot$, the nebular continuum and line emission model \citep{2011MNRAS.415.2920I}, and the dust thermal emission with a modified black-body function.
The dust temperature is assumed to be 30~K, 50~K, or 80~K to account for possibilities of dusty interlopers with various temperatures, and the dust emissivity index is fixed at $\beta_{\rm dust}=2.0$.
The effect of the cosmic microwave background on the dust emission \citep{2013ApJ...766...13D} is also taken into account.

The considered fitting parameters are as follows; 
the SFH time-scale is $\tau_{\rm SFH}=0.01$, $0.03$, $0.06$, $0.1$, $0.3$, $0.6$, $1$, $3$, $6$, and $10$ Gyr (10 cases), 
the metallicity is $Z=0.0001$, $0.0004$, $0.004$, $0.008$, $0.02(=Z_\odot)$, and $0.05$ (6 cases), 
the dust attenuation is $A_{\rm V}=0.01$ to $10$ with 20 logarithmic steps, 
the stellar population age is 7 cases in 1 Myr to 10 Myr, 8 cases in 10 Myr to 100 Myr, 15 cases in 100 Myr to 1 Gyr, and 8 cases in 1 Gyr to 15 Gyr, but limited by the cosmic age at the redshift of interest, 
and the redshift is 0.1 to \redc{20.0} with a 0.1 step \redc{assuming a flat prior}.

\begin{deluxetable}{cccccc}\redc{
\tablecaption{Physical Properties of Our $H$-Dropout Candidates}
\tablehead{
\colhead{Name} & \colhead{$z_\m{phot}$} & \colhead{$M_\m{UV}$} & \colhead{$SFR_\m{UV}$} & \colhead{$\m{log}M_*$}  & \colhead{$A_\m{V}$} \\
\colhead{(1)}& \colhead{(2)}& \colhead{(3)}& \colhead{(4)} &  \colhead{(5)}& \colhead{(6)} }
\startdata
HD1 & $15.2^{+1.2(1.6)}_{-2.1(2.7)}$$^\dagger$ & $-23.3$$^\dagger$ & $110$$^\dagger$ & $\sim9-11$ & $<0.08$\\
HD2 & $12.3^{+0.4(0.7)}_{-0.3(0.7)}$ & $-23.8$ & $170$ & $\sim9.8-11$ & $\lesssim0.8$
\enddata}
\tablecomments{\redc{(1) Name.
(2) The best photometric redshift with $1\sigma$ ($2\sigma$) errors.
(3) Absolute UV magnitude in units of mag.
(4) SFR estimated from the UV magnitude by using Equation (\ref{eq_LUVSFR}) in units of $M_\odot\ \m{yr^{-1}}$.
(5) \& (6) Stellar mass and dust attenuation suggested by the SED fitting in units of $M_\odot$ and mag, respectively. See Section \ref{ss_SEDfit} for details.\\
$^\dagger$$z=13.27$ is suggested by the ALMA observations for HD1 (see Section \ref{ss_ALMA}), consistent with the photometric redshift estimate within $1\sigma$. The absolute UV magnitude and SFR in this table are calculated based on the assumption of $z=13.27$.
}}
\label{tab_sed}
\end{deluxetable}

Figure \ref{fig_SED} shows the results of the SED fitting analyses, \redc{and Table \ref{tab_sed} summarizes the results}.
The best photometric redshifts are always $z>12$ for both HD1 and HD2 thanks to the sharp discontinuity between $H$ and $K_\m{s}(K)$-bands.
The low redshift solutions are found at $z\sim4$ for both objects with larger $\chi^2$ values than the $z>12$ solutions.
These are Balmer break galaxy solutions, and the dust temperature does not affect them because these solutions have very weak or no dust emission.
Another type of possible solutions is dusty H$\alpha$ emitters at $z\sim2$, although these solutions are not supported by the non-detections in the far-infrared and sub-mm bands.
In these solutions, a strong H$\alpha$ line boosts $K_\m{s}(K)$-band and makes $H-K_\m{s}(K)$ color as red as \redc{$z\sim12-16$} galaxies.
In the very high dust temperature case of 80~K, this solution gives a slightly smaller $\chi^2$ than those of the Balmer break ones at $z\sim4$, while the lower, more normal dust temperature cases do not favor this type of the solution.
Moreover, even the 80~K case is significantly less likely compared to the solutions at $z>12$ ($\Delta\chi^2>4)$.

For HD1, the best-fit redshift is $z=15.2$ ($\chi^2=4.7$), which is in fact out of the ALMA [O~{\sc iii}]88$\mu$m scan (Section 3).
The case of $z=13.3$, corresponding to the possible line feature at $z=13.27$, gives $\chi^2=5.4$. 
Since it is \redc{roughly equally likely (within the $1\sigma$ confidence range, see Table \ref{tab_sed})}, we show this case in Figure~\ref{fig_SED}.
The physical properties are not well constrained, except for the dust attenuation that is $A_{\rm V}<0.08$ ($2\sigma$).
The stellar mass ($M_*$) is $(1$--$100)\times10^9~M_\odot$, depending on the stellar age that is not constrained.
When the age is less than $\sim10$ Myr, the stellar mass and star formation rate (SFR) are estimated to be $M_*\sim1\times10^9~M_\odot$ and $SFR\sim10^{2-3}~M_\odot~{\rm yr}^{-1}$, respectively.
For an age of 10--100 Myr ($>100$ Myr), $M_*\sim(1-10)\times10^9~M_\odot$ and $SFR\sim10^{2}~M_\odot~{\rm yr}^{-1}$ ($M_*\sim(10-100)\times10^9~M_\odot$ and $SFR<10^{2}~M_\odot~{\rm yr}^{-1}$) are obtained.
The SFH time-scale also produces dependencies; for $\tau_{\rm SFH}>100$ Myr (a larger value is closer to a constant SFH), we obtain $M_*\sim(1-10)\times10^9~M_\odot$ and $SFR\sim10^{2-3}~M_\odot~{\rm yr}^{-1}$, and for $\tau_{\rm SFH}<100$ Myr, $M_*$ and $SFR$ show larger variations.
The metallicity is not constrained at all.

For HD2, the best-fit redshift is $z=12.3$ ($\chi^2=4.2$) and the $1\sigma$ range ($\Delta\chi^2<1$) is $12.0<z<12.7$.
We find two types of the high redshift solutions.
One is a very young starburst: 
an age less than 10 Myr, 
$M_*\sim7\times10^9~M_\odot$, $SFR\sim10^{3-4}~M_\odot~{\rm yr}^{-1}$, and $A_{\rm V}\sim0.8$.
$\tau_{\rm SFH}$ and metallicity are not constrained.
The other case is a massive and relatively mature galaxy:
an age greater than 100 Myr, 
$M_*\sim1\times10^{11}~M_\odot$, $SFR<10^{2}~M_\odot~{\rm yr}^{-1}$, $A_{\rm V}<0.5$, and $\tau_{\rm SFH}<60$ Myr.
The metallicity is not constrained.

Although they are statistically less likely given the larger $\chi^2$ values, the possible Balmer break solutions are as follows.
For HD1, we obtain $z\sim3.9$, age of 0.3--1 Gyr, $M_*\sim(6-10)\times10^9~M_\odot$, $SFR<0.1~M_\odot~{\rm yr}^{-1}$, $A_{\rm V}<0.5$, $\tau_{\rm SFH}<0.1$ Gyr, and $Z>0.004$.
For HD2, we obtain $z\sim3.5$, age of 0.4--0.7 Gyr, $M_*\sim1\times10^{10}~M_\odot$, $SFR\sim0~M_\odot~{\rm yr}^{-1}$, $A_{\rm V}<0.1$, $\tau_{\rm SFH}<0.03$ Gyr, and $Z>0.02$.
These stellar masses of $\sim10^{10}\ M_\odot$ are $\sim10$ times smaller than known passive galaxies at $z\sim4$ \citep{2017Natur.544...71G,2019ApJ...885L..34T,2020ApJ...889...93V}.
Therefore, even these cases are also interesting to be examined further spectroscopically in future.

\section{Luminosity Function and SFR Density}\label{ss_UVLF_SFRD}

\subsection{Selection Completeness}\label{ss_completeness}

\begin{figure}
\centering
\includegraphics[width=0.9\hsize, bb=4 6 350 500]{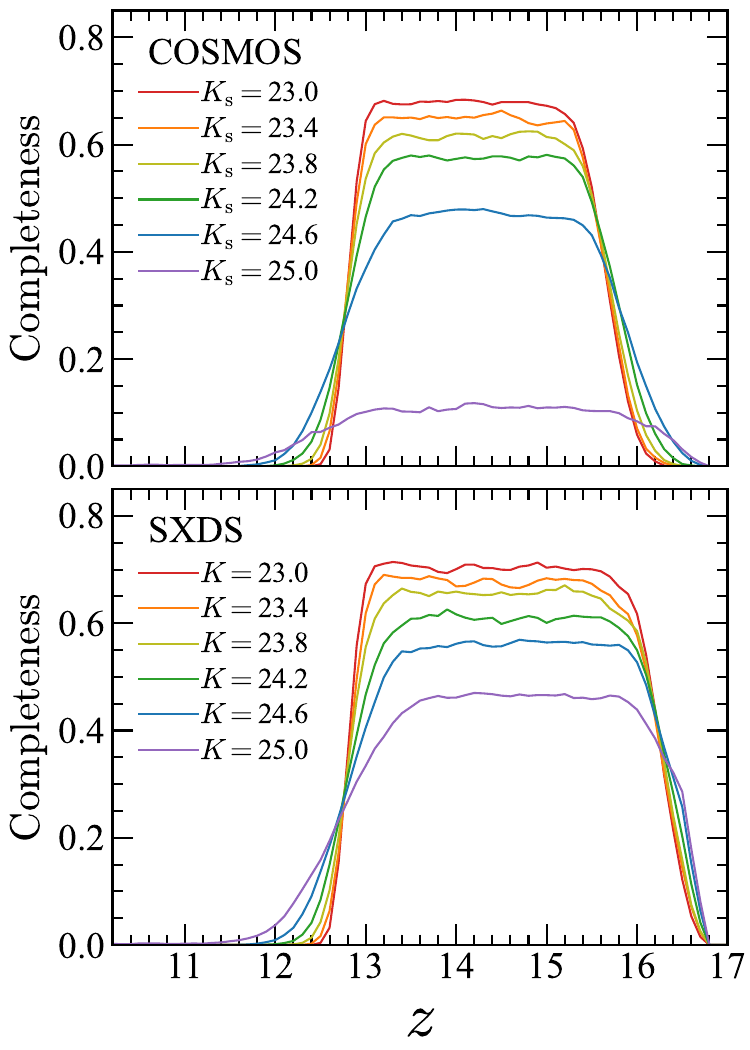}
\caption{Completeness estimated in our Monte Carlo simulations.
The top and bottom panels are results for the COSMOS and SXDS fields, respectively.
The red, orange, yellow, green, blue, and purple lines show the completeness for $K_\m{s} (K)=23.0$, $23.4$, $23.8$, $24.2$, $24.6$, and $25.0$ mag sources, respectively.
}
\label{fig_completeness}
\end{figure}

To derive the rest-frame UV luminosity function of the \redc{$z\sim12-16$} galaxies, we estimate the selection completeness by conducting Monte Carlo simulations.
We first make mock SEDs of galaxies at $9.0<z<19.0$ (steps of 0.1) with UV spectral slopes of $-3.0<\beta_\m{UV}<-1.0$ (steps of 0.1). 
The IGM attenuation is taken into account by using a prescription of \citet{2014MNRAS.442.1805I}, resulting almost zero flux densities at the wavelength bluer than the Ly$\alpha$ break.
We then calculate fluxes in each band by integrating mock SEDs through our 10 filters ($grizyJHK_\m{s}(K)[3.6][4.5]$), and scale to have apparent magnitudes of $23.0-25.0$ mag in the $K_\m{s}(K)$-band, whose central wavelength corresponds to $\sim1500\ \m{\AA}$ at $z\sim13$.
We then perturb the calculated fluxes by adding photometric scatters based on a Gaussian distribution with a standard deviation equal to the flux uncertainties in each band.
We generate 1000 mock galaxies at each redshift with UV spectral slopes following a Gaussian distribution with a mean of $\beta_\m{UV}=-2.0$ and a scatter of $\Delta\beta_\m{UV}=0.2$ \citep{2014MNRAS.440.3714R,2020MNRAS.493.2059B}.
Finally, we select \redc{$z\sim12-16$} galaxy candidates with the same color selection criteria, and calculate the selection completeness as a function of the $K_\m{s}(K)$-band magnitude and redshift, $C(m,z)$, averaged over the UV spectral slope.
Figure \ref{fig_completeness} shows the calculated selection completeness in the COSMOS and SXDS fields.
Our selection criteria can select sources at $12\lesssim z\lesssim16$.
The mean redshift from the simulation is $z=14.3$ and $14.6$ in the COSMOS and SXDS fields, respectively, but in this paper we adopt $z=12.8$ ($z\sim13$) that is the average of the nominal redshifts for HD1 and HD2, as the mean redshift of our $H$-dropout sample.
The selection completeness is $\sim70\%$ even for very bright (23.0 mag) galaxies because our color criteria are very strict in order to remove foreground interlopers (see Figure \ref{fig_2color}) and miss some intrinsically-red ($\beta_\m{UV}\gtrsim-1.8$) $z\gtrsim12$ galaxies.

Based on the results of these selection completeness simulations, we estimate the survey volume per unit area as a function of the apparent magnitude \citep{1999ApJ...519....1S},
\begin{equation}
V_{\rm eff} (m) = \int C(m,z) \frac{dV(z)}{dz} dz, 
\end{equation}
where $C(m,z)$ is the selection completeness, i.e., the probability that a galaxy with an apparent magnitude $m$ at redshift $z$ is detected and satisfies the selection criteria, and $dV(z)/dz$ is the differential comoving volume as a function of redshift.

\subsection{Contamination}\label{ss_cont}

The space number density of the $z\sim13$ galaxies that are corrected for incompleteness and contamination is calculated with the following equation:
\begin{equation}
\psi(m) = \left[1-f_\m{cont}\right]\frac{n_{\rm raw}(m)}{V_{\rm eff}(m)},
\end{equation}
where $n_{\rm raw}(m)$ is the surface number density of selected galaxies in an apparent magnitude bin of $m$, and $f_{\rm cont}$ is a contamination fraction.
We estimate the contamination fraction of foreground sources by conducting Monte Carlo simulations.
As discussed in Section \ref{ss_SEDfit}, the most likely contaminants are $z\sim4$ passive galaxies whose Balmer breaks mimic the Lyman break at $z\sim13$.
Stellar contaminations are not expected to be dominant, given observed colors of stellar sources (Figure \ref{fig_2color}).

To investigate contamination from various $z\sim4$ passive galaxies, we prepare three types of mock SEDs at $3\leq z\leq 5$ based on 1) a classic spectrum of elliptical galaxies in \citet{1980ApJS...43..393C}, 2) model spectra with color distributions similar to real passive galaxies, and 3) the $z\sim4$ solutions from the SED fittings.
In case 1, we use a spectrum of old elliptical galaxies in \citet{1980ApJS...43..393C} as an input SED.
In case 2, we first generate model spectra of galaxies by using {PANHIT} assuming a delayed-$\tau$ star formation history, $\tau=0.01$, $0.03$, $0.1$, $0.3$, and $1\ \m{Gyr}$, the stellar age of $0.01-1.3\ \m{Gyr}$, metallicity of $Z=0.0001$, $0.0004$, $0.004$, $0.008$, $0.02$, and $0.05$, and dust attenuation of $E(B-V)=0-1$ (steps of 0.05).
Then we calculate rest-frame $NUV-r$ and $r-J$ colors of the models, and compare these colors with those of observed passive galaxies in \citet{2017A&A...605A..70D}.
By selecting galaxies whose colors are consistent with the observed passive galaxies, we construct a set of passive galaxy SEDs that have realistic color distributions.
In case 3, we use a spectrum of the $z\sim4$ passive galaxy solution in the SED fitting in Section \ref{ss_SEDfit}.
Since in this case we assume that all of the passive galaxies have the same SED as the $z\sim4$ solution, this case provides the most conservative estimate for the contamination fraction (i.e., the highest contamination fraction).

We then make mock SEDs redshifted to $3\leq z \leq 5$ (steps of 0.1) from the three types of the SEDs, calculate fluxes in each band, scale to have stellar masses of $10^9\leq M_*/M_\odot\leq 10^{11}$ (steps of 0.1 dex), and perturb the calculated fluxes by adding photometric scatters in the same manner as Section \ref{ss_completeness}.
We generate $\sim1000$ mock galaxies at each redshift and stellar mass bin, and calculate the fraction of passive galaxies that satisfy our selection criteria in each bin.
Finally, by integrating the product of the stellar mass function of passive galaxies in \citet{2017A&A...605A..70D} and the fraction of passive galaxies satisfying our selection criteria over the redshift and stellar mass, we calculate the number of passive galaxies at $z\sim4$ that are expected to be in our $z\sim13$ galaxy sample.

The expected numbers of passive galaxies in our sample are $N_\m{cont}=0.00$, $0.12$, $1.36$ in cases of 1), 2), and 3), respectively.
We estimate the contamination fraction $f_\m{cont}$ by dividing the expected number of passive galaxies by the number of our $z\sim13$ candidates.
The estimated contamination fractions are small in cases of 1 and 2 ($f_\m{cont}\sim0\%$ and $6\%$, respectively), and $f_\m{cont}\sim70\%$ in case 3, where we assume all of the passive galaxies have the same SED as the $z\sim4$ solution in the SED fitting as the extremely conservative case.
Although the realistic simulation with the observed color distributions (i.e., case 2) indicates the very low contamination fraction, we adopt this very conservative estimate from case 3 for the UV luminosity function calculation.
\redc{Note that even if we assume this conservative estimate as the prior, the $z>12$ solutions for HD1 and HD2 in the SED fitting are still more likely than the $z\sim4$ solutions that give larger $\chi^2$ values, as long as the true number density of $z\sim13$ galaxies is $\gtrsim10^{-8}\ \m{Mpc^{-3}}$ (comparable with our estimate in Section \ref{ss_UVLF}).
On the other hand, if the true number density is $\sim10^{-11}\ \m{Mpc^{-3}}$ at $z\sim13$ (comparable with model predictions in Section \ref{ss_model}), the $z\sim4$ solutions are more likely due to the higher number density of $z\sim4$ passive galaxies compared to that of $z\sim13$ galaxies.}

\subsection{UV Luminosity Function}\label{ss_UVLF}

We convert the number density of $z\sim13$ galaxies as a function of apparent magnitude, $\psi(m)$, into the UV luminosity functions, $\Phi[M_{\rm UV}(m)]$, which is the number densities of galaxies as a function of rest-frame UV absolute magnitude.
We calculate the absolute UV magnitudes of galaxies from their apparent magnitudes in the $K_\m{s}(K)$-band, whose central wavelength corresponds to $\sim1500\ \m{\AA}$ at $z\sim13$, assuming a flat rest-frame UV continuum, i.e., constant $f_\nu$, suggested by the SEDs of our galaxies.
The $1\sigma$ uncertainty is calculated by taking into account the Poisson confidence limit \citep{1986ApJ...303..336G} on the expected number of galaxies at $z\sim13$ in our sample (\redc{$N=2\times(1-f_\m{cont})\sim1$}). 

\begin{figure}
\centering
\includegraphics[width=0.99\hsize, bb=7 9 430 358]{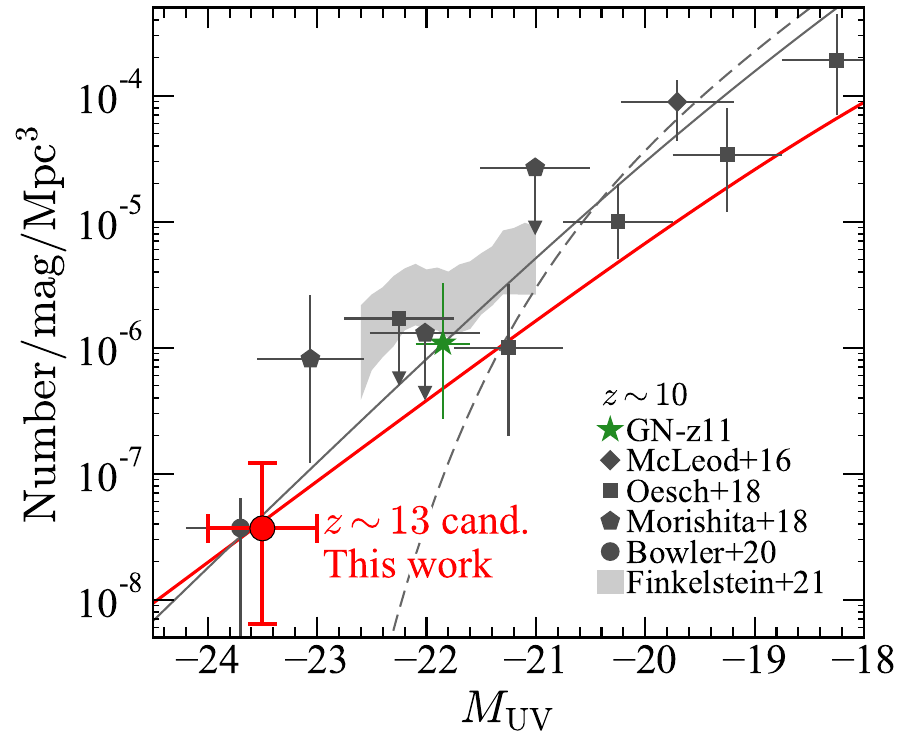}
\caption{
Rest-frame UV luminosity functions at $z\sim13$ and $z\sim10$.
The red circle shows the number density of our $z\sim13$ galaxy candidates.
The black symbols and the gray shaded region are measurements at $z\sim10$ from the literature (diamond: \citealt{2016MNRAS.459.3812M}, square: \citealt{2018ApJ...855..105O}, pentagon: \citealt{2018ApJ...867..150M}, circle: \citealt{2020MNRAS.493.2059B}, shade: \citealt{2021arXiv210613813F}). 
The green star is the number density of GN-z11 (see text).
Note that the data point of \citet{2020MNRAS.493.2059B} (GN-z11) is horizontally (vertically) offset by $-0.2$ mag (+0.03 dex) for clarity.
The gray dashed line is the Schechter function fit \citep{2016ApJ...830...67B}, whereas the gray and red solid lines are the double power-law functions at $z\sim10$ and $13$, respectively, whose parameters are determined by the extrapolation from lower redshifts in \citet{2020MNRAS.493.2059B}.
}
\label{fig_UVLF}
\end{figure}

\begin{figure}
\centering
\includegraphics[width=0.99\hsize, bb=7 9 430 358]{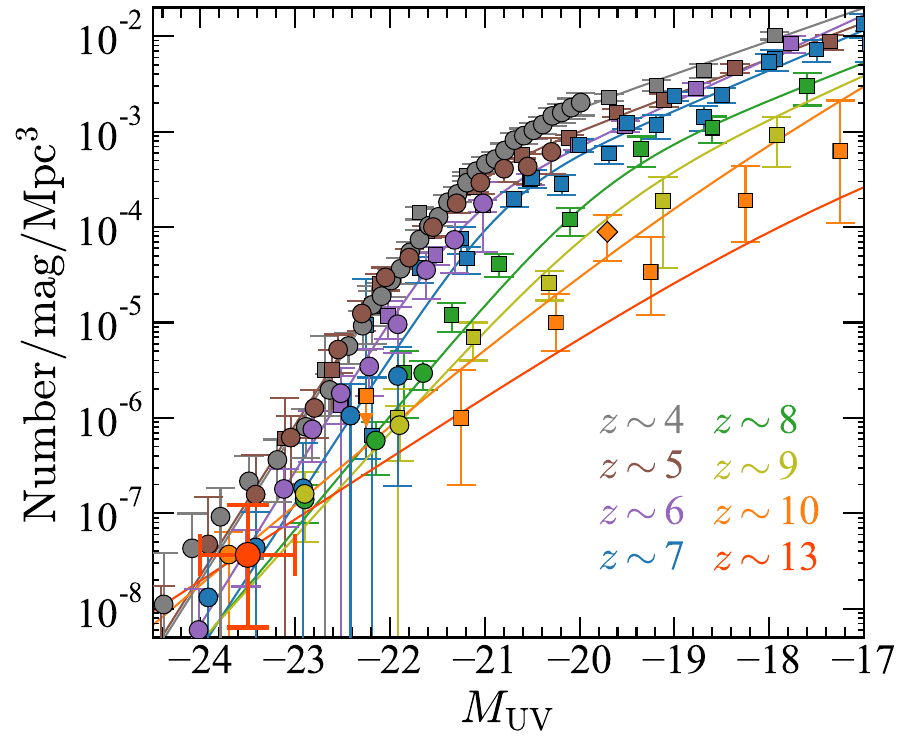}
\caption{
Evolution of the rest-frame UV luminosity functions from $z\sim4$ to $z\sim13$.
The red circle shows the number density of our $z\sim13$ galaxy candidates, and the grey, brown, purple, blue, green, yellow, and orange symbols show results at $z\sim4$, $5$, $6$, $7$, $8$, $9$, and $10$, respectively.
The circles at $z\sim4-7$ are galaxy number densities from \citet{2021arXiv210801090H}, and those at $z\sim8-10$ are taken from \citet{2020MNRAS.493.2059B}.
The squares show results taken from \citet{2021AJ....162...47B} and \citet{2018ApJ...855..105O} at $z\sim4-9$ and $z\sim10$, respectively.
The diamond is a result in \citet{2016MNRAS.459.3812M}.
The lines show double power-law functions in \citet{2021arXiv210801090H} at $z\sim4-7$ and \citet{2020MNRAS.493.2059B} at $z\sim8-13$.
Note that the data point of \citet{2020MNRAS.493.2059B} at $z\sim10$ is horizontally offset by $-0.2$ mag for clarity.
}
\label{fig_UVLF_evol}
\end{figure}

Figure \ref{fig_UVLF} shows the calculated UV luminosity function at $z\sim13$.
The number density of our $z\sim13$ galaxies is $(3.7^{+8.4}_{-3.0})\times10^{-8}\ \m{Mpc^{-3}\ mag^{-1}}$ at $M_\m{UV}=-23.5$ mag.
This number density is comparable to that of bright galaxies at $z\sim10$ in \citet{2020MNRAS.493.2059B}, which is supported by the little evolution of the abundance of bright galaxies found by previous studies at $z=4-10$ \citep{2020MNRAS.493.2059B,2021arXiv210801090H}.
Indeed as shown in Figure \ref{fig_UVLF_evol}, the number density of bright ($M_\m{UV}<-23\ \m{mag}$) galaxies do not show significant redshift evolution from $z\sim4$ to $z\sim13$.
In Figure \ref{fig_UVLF}, we also plot the number density of $z\sim10$ galaxies estimated from GN-z11, $(1.0^{+2.2}_{-0.8})\times10^{-6}\ \m{Mpc^{-3}}$.\footnote{This number density is lower than that in \citet{2020MNRAS.493.2059B}, because we adopt the number density estimate of \citet[][]{2018ApJ...855..105O}. The UV magnitude of GN-z11 is estimated to be $-22.1$ and $-21.6$ mag in \citet[][]{2016ApJ...819..129O} and \citet[][]{2018ApJ...855..105O}, respectively. We adopt their average value, $-21.85\pm0.25$ mag, which is consistent with the recent estimate by \citet{2021arXiv211105351T}.}
These results and the spectroscopic confirmation of GN-z11 by \citet{2016ApJ...819..129O} and \citet{2021NatAs...5..256J} indicate that the bright end of the luminosity function at high redshift cannot be explained by the Schechter function with the exponential cutoff, and is more consistent with the double power-law function.
Indeed, the number density of our $z\sim13$ galaxies is consistent with the double power-law function with $M_\m{UV}^*=-17.6$ mag, $\phi^*=1.0\times10^{-4}\ \m{Mpc^{-3}}$, $\alpha=-1.8$, and $\beta=-2.6$ (Figure \ref{fig_UVLF}), which are derived by extrapolating the redshift evolution of the parameters in \citet{2020MNRAS.493.2059B} to $z=13$.\footnote{\redc{The extrapolated double power-law luminosity function from \citet{2020MNRAS.493.2059B} predicts a $\sim2$ times higher number density at $z\sim11-12$ than those at $z\sim10$ and $13$ in the magnitude regime of $M_\m{UV}\simeq-23.5$ mag, but still consistent with our $z\sim13$ estimate within the errors. The detail of such a redshift evolution is beyond the scope of this paper.}}
Although spectroscopic confirmation is needed, these results indicate that upcoming surveys will detect a number of galaxies at $z>10$, which will be discussed later in Section \ref{ss_future}.

\begin{figure*}
\centering
\begin{minipage}{0.49\hsize}
\begin{center}
\includegraphics[width=0.9\hsize, bb=17 11 414 354]{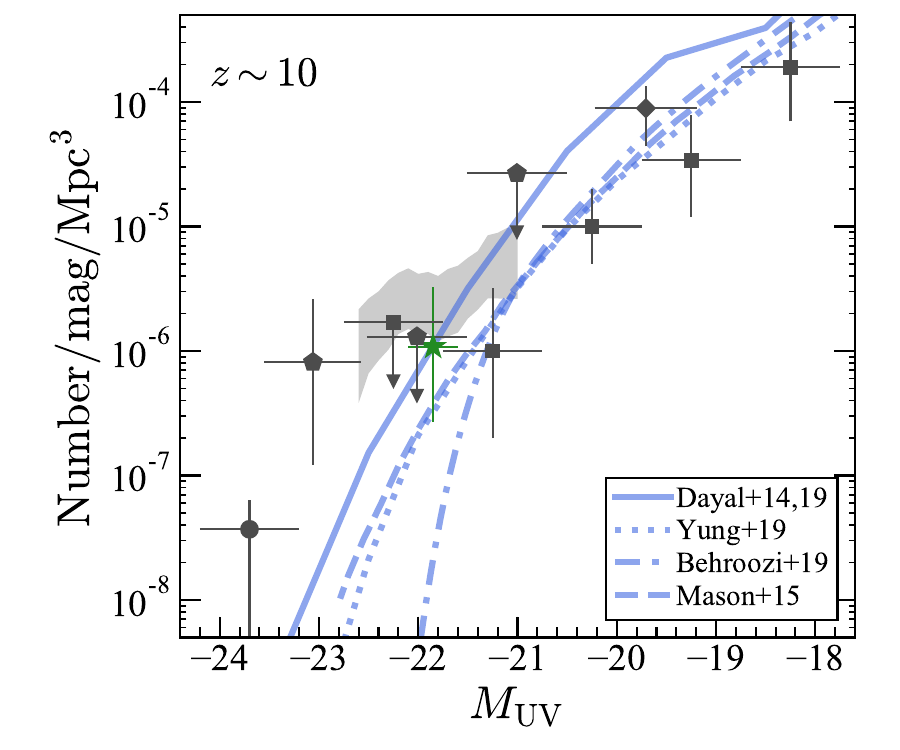}
\end{center}
\end{minipage}
\begin{minipage}{0.49\hsize}
\begin{center}
\includegraphics[width=0.9\hsize, bb=17 11 414 354]{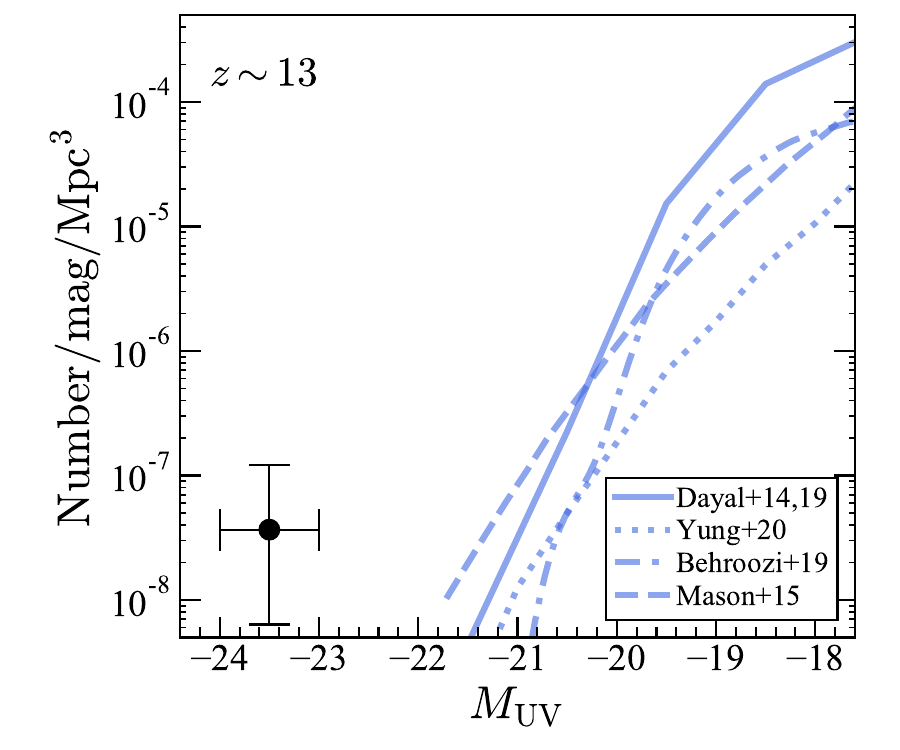}
\end{center}
\end{minipage}
\caption{Comparison with predictions from theoretical and empirical models at $z\sim10$ (left) and $z\sim13$ (right).
The black symbols and the gray shaded region are measurements at $z\sim10$ from the literature (symbols are the same as in Figure \ref{fig_UVLF}) and at $z\sim13$ from this study.
The blue lines show predictions from models (solid: \citealt{2014MNRAS.445.2545D,2019MNRAS.486.2336D}, dotted: \citealt{2019MNRAS.483.2983Y,2020MNRAS.496.4574Y}, dot-dashed: \citealt{2019MNRAS.488.3143B,2020MNRAS.499.5702B}, dashed: \citealt{2015ApJ...813...21M}).
}
\label{fig_UVLF_model}
\end{figure*}

\subsection{Comparison with Models}\label{ss_model}

Both theoretical and empirical models predict the UV luminosity function of galaxies at $z>10$ (e.g., \citealt{2014MNRAS.445.2545D,2019MNRAS.486.2336D}, \citealt{2015ApJ...813...21M}, \citealt{2018ApJ...868...92T}, \citealt{2019MNRAS.488.3143B,2020MNRAS.499.5702B}, \citealt{2019MNRAS.483.2983Y,2020MNRAS.496.4574Y}, see also \citealt{2021MNRAS.503.3698H}).
We compare the number densities at $z\sim10$ and $13$ with predictions from these models in Figure \ref{fig_UVLF_model}.
At $z\sim10$, the predictions roughly agree with the observed number densities for relatively faint galaxies ($M_\m{UV}\gtrsim-21$ mag), but the models underestimate the number densities of bright galaxies ($M_\m{UV}\lesssim-22$ mag) albeit with large uncertainties in the observations.
Similarly at $z\sim13$, the models cannot reproduce the observed number density of our $z\sim13$ galaxy candidates.
These discrepancies indicate that the current models do not account for the rapid mass growth within the short physical time since the Big Bang.

There are several possible physical processes to reconcile these discrepancies between the models and the observations at $z\sim10-13$.
As discussed in \citet{2021arXiv210801090H}, less efficient mass quenching and/or lower dust obscuration than assumed in the models can explain the existence of these UV-bright galaxies.
AGN activity may also boost the UV luminosity in these galaxies.
Previous studies indicate that the AGN fraction starts to increase at $M_\m{UV}\simeq-22\ \m{mag}$ (\citealt{2018PASJ...70S..10O,2018ApJ...863...63S,2020MNRAS.494.1771A,2021arXiv210801090H}, see also \citealt{2021arXiv211103105P}).
If we assume that the UV luminosities of HD1 and HD2 are solely powered by black holes, the inferred black hole masses are $\sim10^8\ \m{M_\odot}$, assuming accretion at the Eddington rate \citep{2022arXiv220100823P}, in accordance with expectations for high redshift quasars (see, e.g., \citealt{2000ApJ...531...42H} and \citealt{2010AJ....140..546W}).
In addition, note that a $\sim 10^8\ M_\odot$ black hole at $z\sim12$ could be the progenitor of $z\sim7$ quasars, as the growth time to reach a mass of $10^{9-10}\ M_\odot$, typical of $z>6$ quasars detected thus far, is shorter than the cosmic time between $z=12-13$ and $z=7$, for an Eddington-limited accretion.
It is also possible that the observed bright source at $z\sim10-13$ are galaxies in a short-time starburst phase that is not captured in the models whose outputs are averaged over a time interval \citep[see also][]{2013MNRAS.434.1486D}.
Finally, a top-heavier IMF would explain the discrepancies by producing more UV photons at the same stellar mass.
It is possible that these bright galaxies (especially HD1 and HD2) are merging systems that are not resolved in the ground-based images.
However, even if they are major mergers, the UV luminosity will decrease only by a factor of a few, which would not explain the discrepancy at $z\sim13$ (see also discussions in \citealt{2021arXiv210801090H} and \citealt{2021arXiv210603728S}).
In any case, if these bright $z\sim10-13$ galaxies are spectroscopically confirmed, the discrepancies will motivate the exploration of new physical processes that are responsible for driving the formation of these bright galaxies in the early universe.

\subsection{Cosmic SFR Density}

We calculate the cosmic SFR density at $z\sim13$ by integrating the rest-frame UV luminosity function.
We use a double power-law luminosity function at $z=13$ with $M_\m{UV}^*=-17.6$ mag, $\phi^*=1.0\times10^{-4}\ \m{Mpc^{-3}}$ $\alpha=-1.8$, and $\beta=-2.6$, which is consistent with our number density measurement (see Section \ref{ss_UVLF}).
We obtain the UV luminosity density by integrating the luminosity function down to $-17$ mag as previous studies \citep[e.g.,][]{2015ApJ...803...34B,2020ApJ...902..112B,2015ApJ...810...71F,2018ApJ...855..105O}.
We then convert the UV luminosity density to the SFR density by using the calibration used in \citet{2014ARA&A..52..415M} with the \citet{1955ApJ...121..161S} IMF:
\begin{equation}
SFR_\m{UV}\ (M_\odot\ \m{yr}^{-1})=1.15\times10^{-28} L_\m{UV}\ (\m{erg\ s^{-1}\ Hz^{-1}}).\label{eq_LUVSFR}
\end{equation}
This SFR density estimation is true only if the rest-frame UV galaxy identification is complete with respect to the all galaxy populations at $z\sim13$ (but see \citealt{2020A&A...643A...4F}).
The uncertainty of the SFR density is scaled from that of the number density measurement. 

\begin{figure}
\centering
\includegraphics[width=0.99\hsize, bb=9 7 389 283]{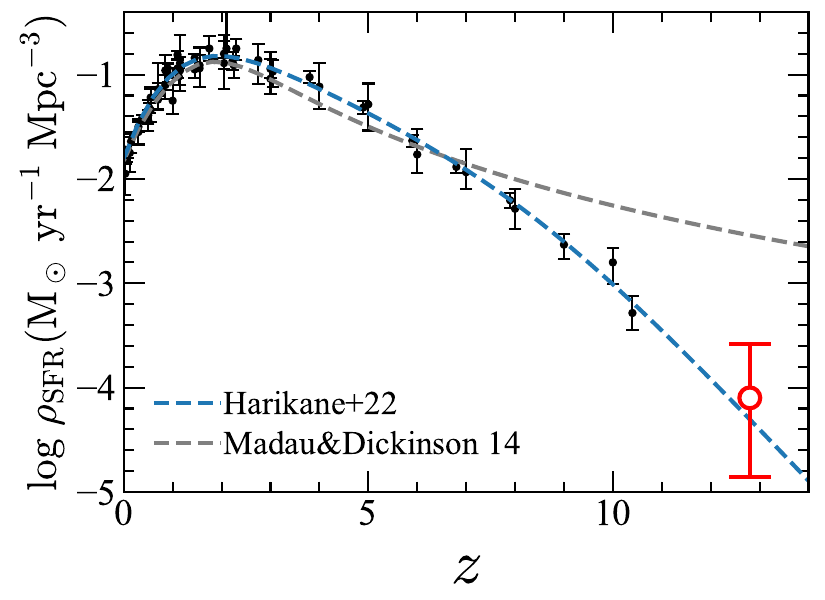}
\caption{
Evolution of the cosmic SFR density.
The red circle is our result at $z\sim13$ estimated by integrating the double power-law luminosity function down to $-17$ mag.
The black circles are observed cosmic SFR densities taken from \citet{2014ARA&A..52..415M}, \citet{2015ApJ...810...71F}, \citet{2016MNRAS.459.3812M}, and \citet{2020ApJ...902..112B}.
The blue and gray dashed curves represent the fits in \citet[][their Equation (60)]{2021arXiv210801090H} and \citet[][extrapolated at $z>8$]{2014ARA&A..52..415M}, respectively.
All results are converted to use the \citet{1955ApJ...121..161S} IMF (Equation (\ref{eq_LUVSFR})).
}
\label{fig_cSFR}
\end{figure}

The estimated SFR density is $\rho_\m{SFR}=(8.0^{+18.4}_{-6.6})\times10^{-5}\ \m{M_\odot\ yr^{-1}\ Mpc^{-3}}$ at $z=12.8$.
We compare the SFR density with previous results in Figure \ref{fig_cSFR}.
The estimated SFR density at $z\sim13$ is consistent with the fitting function in \citet{2021arXiv210801090H}, which is calibrated with recent observations at $z>4$ showing a more rapid decline ($\propto10^{-0.5(1+z)}$) than the extrapolation of the fitting function in \citet{2014ARA&A..52..415M} at $z>10$.
\redc{Furthermore, if we divide our sample into one galaxy at $z\sim12$ and another at $z\sim13$, given the low completeness at $z=12.3$ (see Figure \ref{fig_completeness}), the estimated SFR density would show a decrease from $z\sim12$ to $13$, consistent with \citet{2021arXiv210801090H}.}
The estimate is also comparable to the edge of the range of the SFR density at $z\gtrsim13$ expected from passive galaxies at $z\sim6$ in \citet{2020ApJ...889..137M}.
Note that our estimated SFR density is dominated by relatively faint ($M_\m{UV}\gtrsim-20$ mag) galaxies (see \citealt{2021arXiv210606544R} for discussions at $z\sim8-10$).
In this calculation, we need to assume the shape of the UV luminosity function, because there is no constraint on the number density of faint galaxies at $z\sim13$.
Future {\it JWST} observations will measure the number density of faint $z\sim13$ galaxies and constrain the shape of the luminosity function combined with this work for bright galaxies, allowing a more robust measurement of the SFR density.

\section{Future Prospect}\label{ss_future}

There are several upcoming space-based missions that can search for $z>10$ galaxies such as {\it JWST}, {\it Nancy Grace Roman Space Telescope} (hereafter {\it Roman}), and {\it Galaxy Reionization EXplorer and PLanetary Universe Spectrometer (GREX-PLUS)}.
By taking advantage of the sensitivity of infrared observations in space, these missions are expected to detect galaxies at $z>10$.
In this section, we will discuss future prospects of these space missions based on our result of the \redc{$z\sim12-16$} galaxy search.

We here consider three missions, {\it JWST}, {\it Roman}, and {\it GREX-PLUS}, whose survey parameters are summarized in Table \ref{tab_future}.
Although {\it Euclid} has a remarkable capability to conduct wide-field surveys, it has a limited wavelength coverage to the $H$-band and can only select sources up to $z\sim10$ ($J$-dropout).
We detail the three missions and survey plans below.

\begin{deluxetable*}{ccccccc}
\tablecaption{Expected Number of $z\gtrsim13$ Galaxies Identified in Upcoming Surveys}
\tablehead{
\colhead{Telescope} & \colhead{Survey} & \colhead{$m_{5\sigma}$} & \colhead{$A_\m{survey}$} & \colhead{$N(z\sim13)$} & \colhead{$N(z\sim15)$} & \colhead{$N(z\sim17)$} \\
\colhead{(1)}& \colhead{(2)}& \colhead{(3)}& \colhead{(4)} &  \colhead{(5)}& \colhead{(6)}& \colhead{(7)} }
\startdata
{\it JWST} & JADES/Deep & 30.6/30.2 & 0.013 & $11-0.5$ & $1-0$ & $0-0$\\
 & JADES/Medium & 29.7/29.3 & 0.053 & $15-0.5$ & $1-0$ & $0-0$\\
 & CEERS & 29.0/29.2 & 0.027 & $3-0$ & $0.5-0$ & $0-0$\\
 & \redc{COSMOS-Web} & -/28.1$^*$ & 0.6 & $19-0.5$ & $2-0$ & $0-0$\\
 & PRIMER & $\leq$27.8/$\leq$27.7$^\dagger$ & 0.111 & $5-0$ & $0.5-0$ & $0-0$\\
 & NGDEEP & 30.6/30.7 & 0.0027 & $2-0$ & $1-0$ & $0-0$\\
 & \redc{PANORAMIC} & 28.3/28.3 & 0.4 & $17-0.5$ & $2-0$ & $0-0$\\
{\it Roman} & UltraDeep & 29.6/28.6 & 1 & $250-7^\S$ & $5-0$  & \nodata \\
 & Deep & 27.5/27.2 & 20 & $270-3^\S$  & $9-0$ & \nodata\\
 & HLS & 26.7/- & 2000 & $8441-33^\S$  & \nodata  & \nodata\\
{\it GREX-PLUS} & UltraDeep & 27.7 & 1 & $18-0$  & $1-0$ & $0-0$\\
 & Deep & 27.0 & 40 & $262-2$  & $17-0$ & $1-0$\\
 & Medium & 26.0 & 200 & $300-0$  & $17-0$ & $1-0$\\
 & Wide & 24.5 & 2000 & $322-0$  & $14-0$ & $0.5-0$
\enddata
\tablecomments{(1) Telescope.
(2) Planned survey.
(3) $5\sigma$ depth in the AB magnitude in the rest-frame UV band.
In {\it JWST}, we quote depths of the F200W and F227W bands for the $z\sim13$ and $z\sim15-17$ galaxy selections, respectively.
In {\it Roman}, depths of the F184 and K213 bands are quoted for the $z\sim13$ and $z\sim15$ galaxy selections, respectively.
In {\it GREX-PLUS}, we quote depths in the F232 band for the $z\sim13-17$ galaxy selection.
These depths are for point sources for simplicity and the actual depths for high redshift galaxies would be slightly shallower, except in PRIMER and NGDEEP, where resolved sources with typical sizes of high redshift galaxies are assumed.
(4) Survey area in deg$^2$.
(5)-(7) Expected number of galaxies at $z\sim13$, $15$, and $17$ identified in the survey assuming $\Delta z=1$. Two values indicate the numbers in case A and case B (see text for details). 
Note that it may be difficult to select $z\sim17$ sources with {\it Roman} due to the lack of observational bands redder than the K213 band.
Because the current plan of the {\it Roman} HLS does not include the K213 imaging, we do not calculate the expected number of $z\sim15$ galaxies.\\
\redc{$^*$The COSMOS-Web survey will take NIRCam F115W, F150W, F277W, and F444W images. We use the depth of the F277W band for the $z\sim13-17$ galaxy selections.\\}
$^\dagger$PRIMER is a ``wedding cake" survey that is composed of several surveys with different depths and areas \redc{(27.8/27.7 mag for $400\ \m{arcmin^2}$, 28.3/28.3 mag for $300\ \m{arcmin^2}$, and 28.8/28.8 mag for $33\ \m{arcmin^2}$)}.\\
\redc{$^\S$Although the expected number in the {\it Roman} HLS is larger than the other two {\it Roman} surveys, these three surveys will identify galaxies in different luminosities, and are complementary to each other. Please see Figure \ref{fig_UVLF_future_z13} for luminosity ranges that each survey covers.}
}
\label{tab_future}
\end{deluxetable*}

{\it JWST} is NASA's infrared space telescope that was launched on 2021 December 25th.
Thanks to its large 6.5 m-diameter mirror, the sensitivity of {\it JWST} in the infrared is much higher than previous and current observational facilities.
NIRCam is a near-infrared camera at $0.6-5.0\ \m{\mu m}$ whose FoV is roughly $2\times2.2\times2.2\ \m{arcmin^2}$ \citep{2005SPIE.5904....1R}.
In this paper we consider the following six surveys using NIRCam; the JWST Advanced Deep Extragalactic Survey (JADES; Guaranteed Time Observation (GTO) program, \citealt{2017jwst.prop.1181E,2017jwst.prop.1180E}), the Cosmic Evolution Early Release Science (CEERS) survey (ERS-1345; \citealt{2017jwst.prop.1345F}), \redc{the COSMOS-Web survey (GO-1727; \citealt{2021jwst.prop.1727K})}, Public Release IMaging for Extragalactic Research (PRIMER; GO-1837; \citealt{2021jwst.prop.1837D}), the Next Generation Deep Extragalactic Exploratory Public (NGDEEP) survey (GO-2079; \citealt{2021jwst.prop.2079F}), and \redc{the Parallel wide-Area Nircam Observations to Reveal And Measure the Invisible Cosmos (PANORAMIC) survey (GO-2514; \citealt{2021jwst.prop.2514W})}.
These surveys will take deep NIRCam imaging data at $\sim1-5\ \m{\mu m}$ in \redc{$\sim10-2000\ \m{arcmin^2}$} survey fields.
Other high redshift galaxy surveys are also planned in Cycle 1.
For example, Through the Looking GLASS (ERS-1324, \citealt{2017jwst.prop.1324T}) and the Ultra-deep NIRCam and NIRSpec Observations Before the Epoch of Reionization (UNCOVER; GO-2561; \citealt{2021jwst.prop.2561L}) will image the gravitational lensing cluster Abell 2744 with NIRCam.
The Webb Medium-Deep Field survey (GTO-1176; \citealt{2017jwst.prop.1176W}) will use 110 hours to observe 13 medium-deep ($28.4-29.4$ mag) fields, including the James Webb Space Telescope North Ecliptic Pole Time-domain Field \citep{2018PASP..130l4001J}.

{\it Roman} is NASA's optical to near-infrared space telescope whose launch is targeted around mid-2020s.
Although the size of the mirror is comparable to that of {\it HST}, {\it Roman} is expected to conduct wide-field surveys in the near-infrared by taking advantage of the wide FoV of its camera ($0.28\ \m{deg^2}$).
The high latitude survey (HLS) will take images in the Y106, J129, H158, and F184-bands over the $\sim2000\ \m{deg^2}$ sky reaching $\sim26.7$ mag in F184. 
Two additional survey concepts are potentially possible; the {\it Roman} ultra-deep field survey (hereafter UltraDeep; \citealt{2019BAAS...51c.550K}), which will take very deep images (reaching $\sim$ 30 mag in the bluer filters, and 29.6 and 28.6 mag in F184 and K213, respectively) in a small area ($\lesssim1\ \m{deg^2}$), and the {\it Roman} cosmic dawn survey (hereafter Deep; see also \citealt{2018AAS...23125817R}), which will conduct a relatively wide and deep survey ($\sim20\ \m{deg^2}$, 27.5 and 27.2 mag in F184 and K213).\footnote{The sensitivities of the survey are calculated based on the tables in the following website:\\ \url{https://roman.gsfc.nasa.gov/science/anticipated_performance_tables.html}}
Both of these two possible surveys would cover a wavelength range up to 2.3 $\mu$m (the K213 filter), allowing us to select $z\sim15$ galaxies with the F184-dropout selection.

{\it GREX-PLUS} is a new 1.2-m class, cryogenic, wide-field infrared space telescope mission concept proposed to ISAS/JAXA for its launch around mid-2030s.
{\it GREX-PLUS} is planned to have a wide-field camera that will efficiently take wide and deep images at $2-10\ \m{\mu m}$, allowing us to select galaxies at $z\sim10-17$.
Four surveys with different depths and areas (UltraDeep, Deep, Medium, and Wide) are now being planned.

\begin{figure*}
\centering
\begin{minipage}{0.49\hsize}
\begin{center}
\includegraphics[width=0.9\hsize, bb=17 11 428 350]{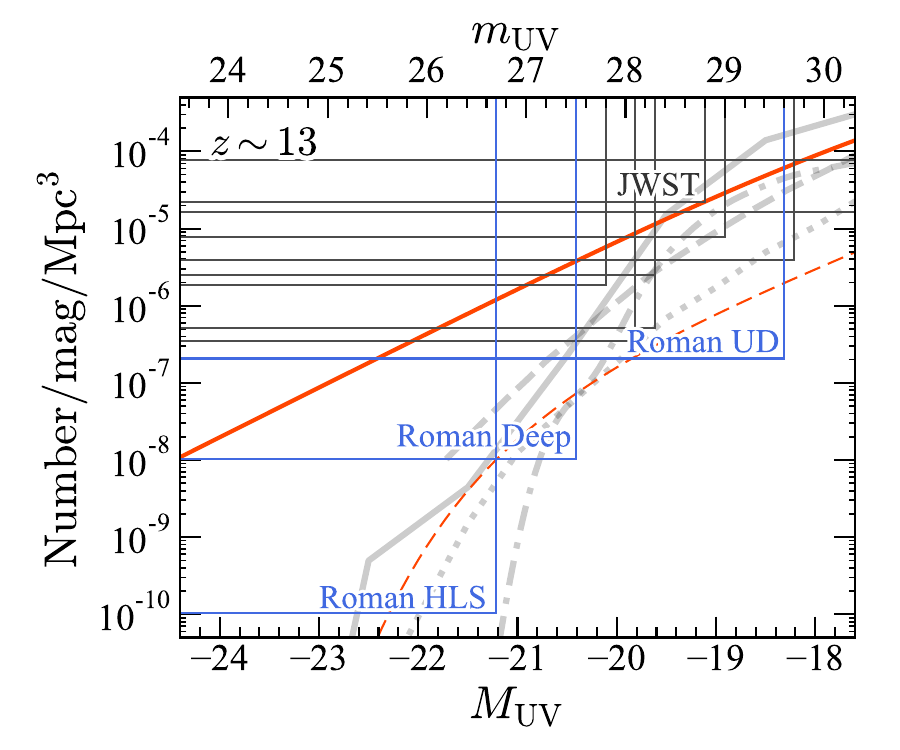}
\end{center}
\end{minipage}
\begin{minipage}{0.49\hsize}
\begin{center}
\includegraphics[width=0.9\hsize, bb=17 11 428 350]{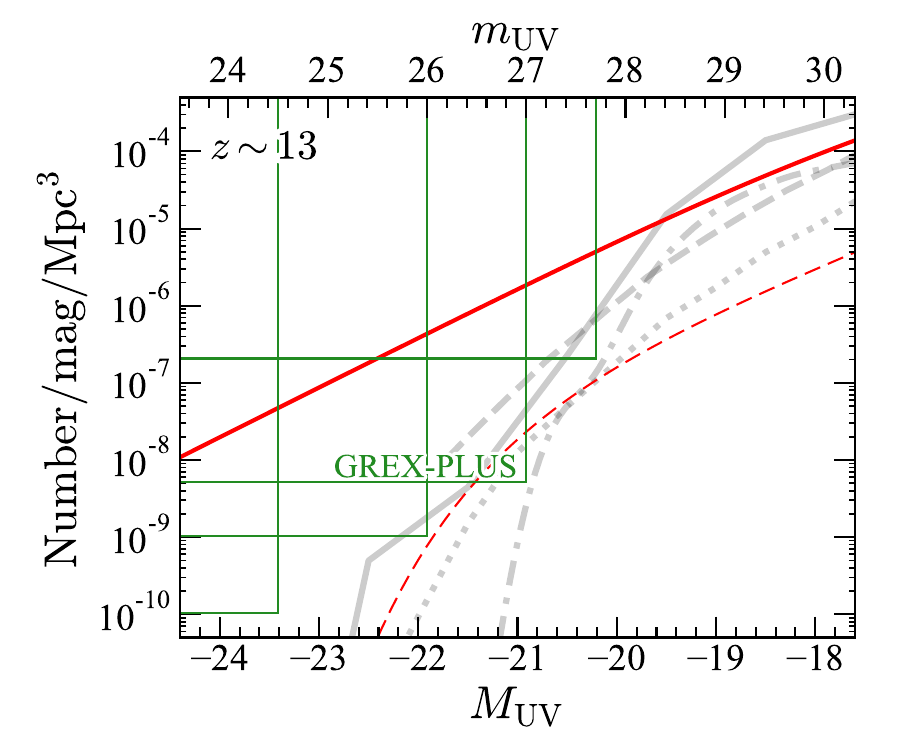}
\end{center}
\end{minipage}
\caption{Predicted UV luminosity functions and depths and volumes of upcoming surveys.
The red solid and dashed lines are the double power-law and Schechter functions (case A and case B in the text), respectively, whose parameters are determined by the extrapolation from lower redshifts (see text for details).
The black, blue ({\it left}) and green ({\it right}) lines indicate expected depths and volumes of upcoming surveys with {\it JWST}, {\it Roman}, and {\it GREX-PLUS}, respectively.
The gray lines show predictions from models (solid: \citealt{2014MNRAS.445.2545D,2019MNRAS.486.2336D}, dotted: \citealt{2019MNRAS.483.2983Y,2020MNRAS.496.4574Y}, dot-dashed: \citealt{2019MNRAS.488.3143B,2020MNRAS.499.5702B}, dashed: \citealt{2015ApJ...813...21M}, same as the right panel of Figure \ref{fig_UVLF_model}).
}
\label{fig_UVLF_future_z13}
\end{figure*}

\begin{figure*}
\centering
\begin{minipage}{0.49\hsize}
\begin{center}
\includegraphics[width=0.9\hsize, bb=17 11 428 350]{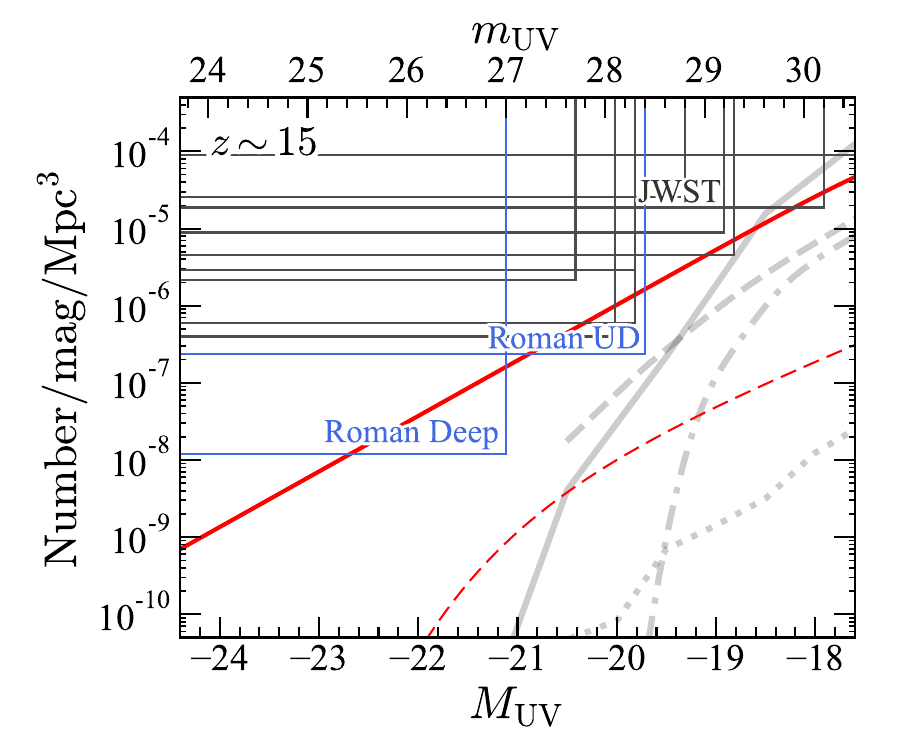}
\end{center}
\end{minipage}
\begin{minipage}{0.49\hsize}
\begin{center}
\includegraphics[width=0.9\hsize, bb=17 11 428 350]{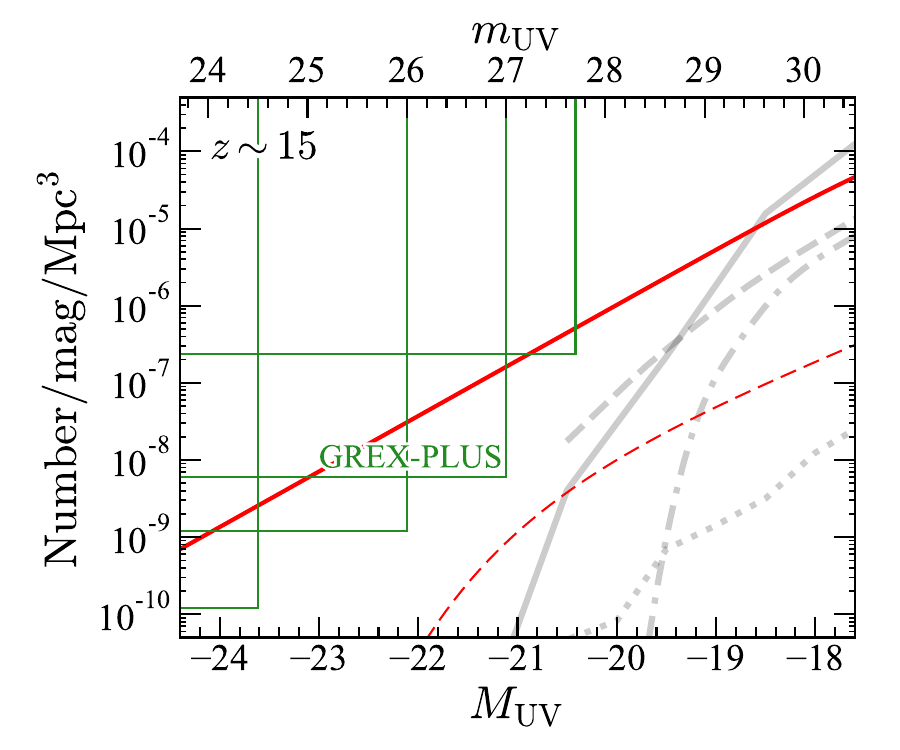}
\end{center}
\end{minipage}
\caption{
Same as Figure \ref{fig_UVLF_future_z13} but for $z\sim15$.
The $6\sigma$ depth is used for {\it Roman} to reduce the false positive rate in the one-band (K213) detection.
}
\label{fig_UVLF_future_z15}
\end{figure*}

For the surveys by these three missions, we calculate the expected number of detected galaxies at $z\sim13$, $15$, and $17$.
For simplicity, we assume the survey volume of $\Delta z=1$ and $100\%$ completeness. 
We calculate the number of galaxies based on two cases of the rest-frame UV luminosity functions (case A and case B).
The one is the double power-law function with a redshift evolution suggested in \citet[][case A]{2020MNRAS.493.2059B}, which is consistent with the number density of our $z\sim13$ galaxies.
The other is the Schechter function with a density evolution, whose parameters are $M_\m{UV}^*=-20.5$ mag, $\phi^*=44.7\times10^{-0.6(1+z)}\ \m{Mpc^{-3}}$, and $\alpha=-2.3$ (case B).
We assume a relatively rapid decrease of the $\phi^*$ parameter compared to the cosmic SFR density evolution at $z>10$ ($\propto10^{-0.5(1+z)}$, \citealt{2021arXiv210801090H}), as a conservative estimate.
These parameters are comparable to measurements for the $z\sim10$ UV luminosity function in \cite{2018ApJ...855..105O}.
These two cases mostly cover the range of model predictions at $z\sim13$ as we show later in Figure \ref{fig_UVLF_future_z13}.
Given that these two cases are somewhat consistent with the model predictions and observations at $z\sim10$ and $13$, we extrapolate these calculations to $z\sim15$ and $17$ for reference.
We integrate the luminosity functions down to the depths of detection bands presented in Table \ref{tab_future}, and calculate the number of detected galaxies at each redshift for each survey.
Note that the $6\sigma$ depth is used for the $z\sim15$ galaxies identified with {\it Roman} to reduce the false positive rate in the one-band (K213) detection, while the $5\sigma$ depth is adopted for the others because multiple bands can be used.

\begin{figure}
\centering
\begin{minipage}{0.99\hsize}
\begin{center}
\includegraphics[width=0.9\hsize, bb=17 11 428 350]{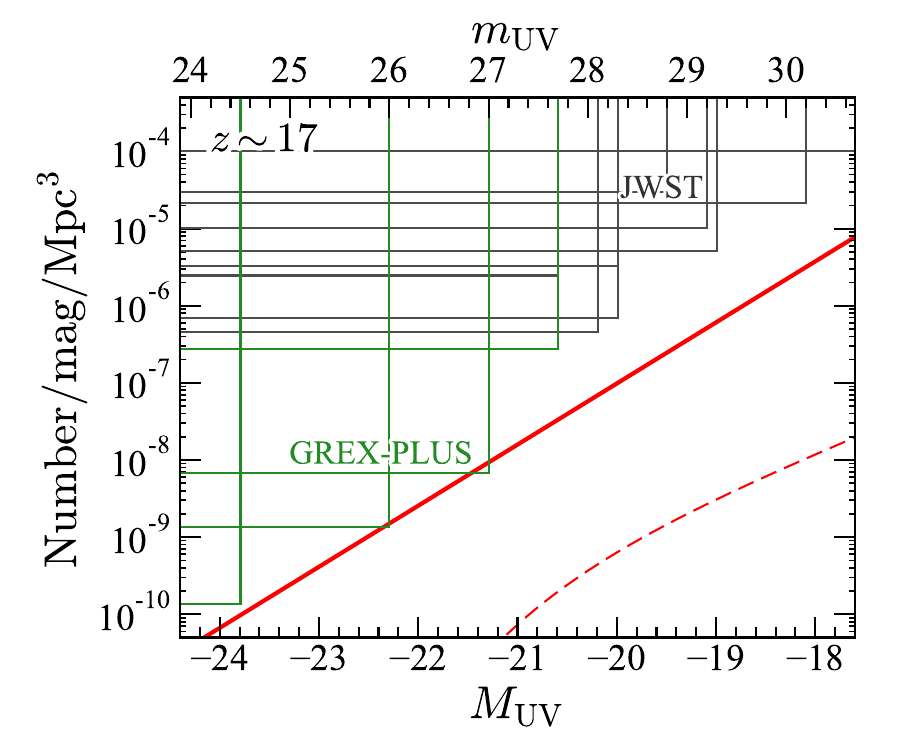}
\end{center}
\end{minipage}
\caption{Same as Figure \ref{fig_UVLF_future_z13} but for $z\sim17$.
We only plot the survey depths and volumes of {\it JWST} and {\it GREX-PLUS}, which cover the wavelength redder than 2.3 $\mu$m (the Ly$\alpha$ break at $z\sim17$).
}
\label{fig_UVLF_future_z17}
\end{figure}

Figures \ref{fig_UVLF_future_z13}, \ref{fig_UVLF_future_z15}, and  \ref{fig_UVLF_future_z17} show the expected UV luminosity functions in case A and case B and the depth and volume of each survey at $z\sim13$, $15$, and $17$, respectively, with predictions from models at $z\sim13$ and $15$.
Table \ref{tab_future} summarizes the expected number of detected galaxies in each survey.
{\it JWST} will identify galaxies up to $z\sim15$ in case A.
In addition, {\it JWST} can conduct deep photometric and spectroscopic observations for relatively bright $z>10$ galaxies identified in surveys with {\it JWST} and other telescopes, which will allow us to investigate physical properties (e.g., systemic redshift, stellar age, metallicity) in detail \citep[e.g.,][]{2021ApJ...910...86R}.
{\it Roman} will detect galaxies up to $z\sim15$ in case A, and identify galaxies at $z\sim13$ even in case B thanks to the wide survey areas.
{\it GREX-PLUS} may be able to push the redshift frontier to $z\sim17$ in case A.
The wide-area surveys with {\it Roman} and {\it GREX-PLUS} can identify luminous $z>10$ galaxies with $\lesssim27\ \m{mag}$.
These galaxies are bright enough to be followed up by spectroscopically with ALMA and {\it JWST} within a reasonable amount of observing time, to investigate the physical conditions of these galaxies in the early universe.

\section{Summary}\label{ss_summary}
In this paper we have presented our search for $H$-dropout LBGs at \redc{$z\sim12-16$}.
We have used the multi-wavelength deep imaging data in the COSMOS and SXDS fields including Subaru/HSC $grizy$, VISTA $JHK_\m{s}$, UKIRT $JHK$, and {\it Spitzer}/IRAC [3.6][4.5] images.
Our major findings are summarized below:
\begin{enumerate}

\item 
After the careful screening of foreground interlopers, we have identified two \redc{$z\sim12-13$} galaxy candidates, HD1 and HD2 (Figure \ref{fig_SED}).
SEDs of these candidates show a sharp discontinuity between $H$ and $K_\m{s}(K)$-bands, non-detections in the $grizyJ$-bands, and a flat continuum up to the [4.5]-band, all of which are consistent with a \redc{$z\sim12-13$} galaxy.
Photometric redshift analyses based on these SEDs indicate that the most likely redshifts are $z>12$ for both HD1 and HD2.

\item 
We calculate the number density of our galaxy candidates whose mean redshift is $z\sim13$ (Figure \ref{fig_UVLF}).
The calculated number density at $z\sim13$ is comparable to that of bright galaxies at $z\sim10$ and consistent with the double power-law luminosity function extrapolated to $z=13$ assuming the redshift evolution in \citet{2020MNRAS.493.2059B}.
These results support little evolution of the abundance of bright galaxies to $z\sim13$ as suggested by previous studies at $z\sim4-10$.
Comparisons with theoretical and empirical models show that these models underestimate the number densities of bright galaxies at $z\sim10-13$, although the uncertainties in observations are large (Figure \ref{fig_UVLF_model}).
The inferred cosmic SFR density is consistent with the rapid decrease at $z>10$ with $\propto10^{-0.5(1+z)}$ suggested by \citet{2021arXiv210801090H} (Figure \ref{fig_cSFR}).

\item 
We conducted ALMA follow-up observations targeting HD1.
The obtained spectrum shows a $\sim4\sigma$ tentative line-like feature at 237.8 GHz that is co-spatial with the rest-frame UV emission, consistent with the {\sc[Oiii]}88$\mu$m emission line at $z=13.27$ (Figure \ref{fig_ALMA}).
Further spectroscopic efforts are needed to confirm the redshifts of HD1 and HD2.

\end{enumerate}

Our results support the possibility that a number of bright galaxies exist at $z>10$.
If the UV luminosity function follows the double-power law function consistent with the number density of the bright galaxies at $z\sim10-13$, upcoming space missions such as {\it JWST}, {\it Roman}, and {\it GREX-PLUS} will detect more than $\sim10000$ galaxies at $z\sim13-15$ (Figures \ref{fig_UVLF_future_z13} and \ref{fig_UVLF_future_z15}), and perhaps one to several at $z\sim17$ (Figure \ref{fig_UVLF_future_z17}), allowing us to observe the first galaxy formation.

\acknowledgments
We thank the anonymous referee for a careful reading and valuable comments that improved the clarity of the paper.
We thank James Rhoads, Sangeeta Malhotra, Masami Ouchi, and the other members in the {\it Roman} cosmic dawn Science Investigation Team (SIT) for helpful discussions on the detectability of $z>10$ galaxies with {\it Roman}.
We are grateful to Caitlin Casey, James Dunlop, Steven Finkelstein, Christina Williams, and Rogier Windhorst for providing the expected depths in their JWST surveys, namely COSMOS-Web, PRIMER, NGDEEP, PANORAMIC, and the Webb Medium-Deep Field survey, respectively.
We thank Takashiro Morishita for bringing an error in Figure 6 in the earlier manuscript to our attention.
This work was partly supported by the joint research program of the Institute for Cosmic Ray Research (ICRR), University of Tokyo, JSPS KAKENHI Grant Numbers 17H06130, 19J01222, 20K22358, and 21K13953, the NAOJ ALMA Scientific Research Grant Codes 2018-09B and 2020-16B, and the Black Hole Initiative at Harvard University, which is funded by grants from the John Templeton Foundation and the Gordon and Betty Moore Foundation.
T.H. was supported by Leading Initiative for Excellent Young Researchers, MEXT, Japan (HJH02007).
P.D. and A. H. acknowledge support from the European Research Council's starting grant ERC StG-717001 (``DELPHI'').
P.D. also acknowledges support from the NWO grant 016.VIDI.189.162 (``ODIN'') and the European Commission's and University of Groningen's CO-FUND Rosalind Franklin program.
A.Y. is supported by an appointment to the NASA Postdoctoral Program (NPP) at NASA Goddard Space Flight Center, administered by Oak Ridge Associated Universities under contract with NASA.
F.P. acknowledges support from a Clay Fellowship administered by the Smithsonian Astrophysical Observatory.

This paper makes use of the following ALMA data: ADS/JAO. ALMA \#2019.A.00015.S .
ALMA is a partnership of ESO (representing its member states), NSF (USA) and NINS (Japan), together with NRC (Canada), MOST, and ASIAA (Taiwan), and KASI (Republic of Korea), in cooperation with the Republic of Chile. The Joint ALMA Observatory is operated by ESO, AUI/NRAO, and NAOJ.

This work is based on data products from observations made with ESO Telescopes at the La Silla Paranal Observatory under ESO programme ID 179.A-2005 and on data products produced by CALET and the Cambridge Astronomy Survey Unit on behalf of the UltraVISTA consortium

The HSC collaboration includes the astronomical communities of Japan and Taiwan, and Princeton University.  The HSC instrumentation and software were developed by the National Astronomical Observatory of Japan (NAOJ), the Kavli Institute for the Physics and Mathematics of the Universe (Kavli IPMU), the University of Tokyo, the High Energy Accelerator Research Organization (KEK), the Academia Sinica Institute for Astronomy and Astrophysics in Taiwan (ASIAA), and Princeton University.  Funding was contributed by the FIRST program from the Japanese Cabinet Office, the Ministry of Education, Culture, Sports, Science and Technology (MEXT), the Japan Society for the Promotion of Science (JSPS), Japan Science and Technology Agency  (JST), the Toray Science  Foundation, NAOJ, Kavli IPMU, KEK, ASIAA, and Princeton University.

This paper makes use of software developed for the Large Synoptic Survey Telescope. We thank the LSST Project for making their code available as free software at  http://dm.lsst.org

This paper is based on data collected at the Subaru Telescope and retrieved from the HSC data archive system, which is operated by Subaru Telescope and Astronomy Data Center (ADC) at NAOJ. Data analysis was in part carried out with the cooperation of Center for Computational Astrophysics (CfCA), NAOJ.

The Pan-STARRS1 Surveys (PS1) and the PS1 public science archive have been made possible through contributions by the Institute for Astronomy, the University of Hawaii, the Pan-STARRS Project Office, the Max Planck Society and its participating institutes, the Max Planck Institute for Astronomy, Heidelberg, and the Max Planck Institute for Extraterrestrial Physics, Garching, The Johns Hopkins University, Durham University, the University of Edinburgh, the Queen’s University Belfast, the Harvard-Smithsonian Center for Astrophysics, the Las Cumbres Observatory Global Telescope Network Incorporated, the National Central University of Taiwan, the Space Telescope Science Institute, the National Aeronautics and Space Administration under grant No. NNX08AR22G issued through the Planetary Science Division of the NASA Science Mission Directorate, the National Science Foundation grant No. AST-1238877, the University of Maryland, Eotvos Lorand University (ELTE), the Los Alamos National Laboratory, and the Gordon and Betty Moore Foundation.

\software{BEAGLE \citep{2016MNRAS.462.1415C}, PAN-HIT (\citealt{2020IAUS..341..285M}), SExtractor \citep{1996A&AS..117..393B}, T-PHOT \citep{2016A&A...595A..97M}}

\bibliographystyle{apj}
\bibliography{apj-jour,reference}

\begin{thebibliography}{}
\expandafter\ifx\csname natexlab\endcsname\relax\def\natexlab#1{#1}\fi

\bibitem[{{Adams} {et~al.}(2020){Adams}, {Bowler}, {Jarvis}, {H{\"a}u{\ss}ler},
  {McLure}, {Bunker}, {Dunlop}, \& {Verma}}]{2020MNRAS.494.1771A}
{Adams}, N.~J., {Bowler}, R.~A.~A., {Jarvis}, M.~J., {et~al.} 2020, \mnras,
  494, 1771

\bibitem[{{Aihara} {et~al.}(2018){Aihara}, {Arimoto}, {Armstrong}, {Arnouts},
  {Bahcall}, {Bickerton}, {Bosch}, {Bundy}, {Capak}, {Chan}, {Chiba}, {Coupon},
  {Egami}, {Enoki}, {Finet}, {Fujimori}, {Fujimoto}, {Furusawa}, {Furusawa},
  {Goto}, {Goulding}, {Greco}, {Greene}, {Gunn}, {Hamana}, {Harikane},
  {Hashimoto}, {Hattori}, {Hayashi}, {Hayashi}, {He{\l}miniak}, {Higuchi},
  {Hikage}, {Ho}, {Hsieh}, {Huang}, {Huang}, {Ikeda}, {Imanishi}, {Inoue},
  {Iwasawa}, {Iwata}, {Jaelani}, {Jian}, {Kamata}, {Karoji}, {Kashikawa},
  {Katayama}, {Kawanomoto}, {Kayo}, {Koda}, {Koike}, {Kojima}, {Komiyama},
  {Konno}, {Koshida}, {Koyama}, {Kusakabe}, {Leauthaud}, {Lee}, {Lin}, {Lin},
  {Lupton}, {Mandelbaum}, {Matsuoka}, {Medezinski}, {Mineo}, {Miyama},
  {Miyatake}, {Miyazaki}, {Momose}, {More}, {More}, {Moritani}, {Moriya},
  {Morokuma}, {Mukae}, {Murata}, {Murayama}, {Nagao}, {Nakata}, {Niida},
  {Niikura}, {Nishizawa}, {Obuchi}, {Oguri}, {Oishi}, {Okabe}, {Okamoto},
  {Okura}, {Ono}, {Onodera}, {Onoue}, {Osato}, {Ouchi}, {Price}, {Pyo}, {Sako},
  {Sawicki}, {Shibuya}, {Shimasaku}, {Shimono}, {Shirasaki}, {Silverman},
  {Simet}, {Speagle}, {Spergel}, {Strauss}, {Sugahara}, {Sugiyama}, {Suto},
  {Suyu}, {Suzuki}, {Tait}, {Takada}, {Takata}, {Tamura}, {Tanaka}, {Tanaka},
  {Tanaka}, {Tanaka}, {Terai}, {Terashima}, {Toba}, {Tominaga}, {Toshikawa},
  {Turner}, {Uchida}, {Uchiyama}, {Umetsu}, {Uraguchi}, {Urata}, {Usuda},
  {Utsumi}, {Wang}, {Wang}, {Wong}, {Yabe}, {Yamada}, {Yamanoi}, {Yasuda},
  {Yeh}, {Yonehara}, \& {Yuma}}]{2018PASJ...70S...4A}
{Aihara}, H., {Arimoto}, N., {Armstrong}, R., {et~al.} 2018, \pasj, 70, S4

\bibitem[{{Aihara} {et~al.}(2019){Aihara}, {AlSayyad}, {Ando}, {Armstrong},
  {Bosch}, {Egami}, {Furusawa}, {Furusawa}, {Goulding}, {Harikane}, {Hikage},
  {Ho}, {Hsieh}, {Huang}, {Ikeda}, {Imanishi}, {Ito}, {Iwata}, {Jaelani},
  {Kakuma}, {Kawana}, {Kikuta}, {Kobayashi}, {Koike}, {Komiyama}, {Li},
  {Liang}, {Lin}, {Luo}, {Lupton}, {Lust}, {MacArthur}, {Matsuoka}, {Mineo},
  {Miyatake}, {Miyazaki}, {More}, {Murata}, {Namiki}, {Nishizawa}, {Oguri},
  {Okabe}, {Okamoto}, {Okura}, {Ono}, {Onodera}, {Onoue}, {Osato}, {Ouchi},
  {Shibuya}, {Strauss}, {Sugiyama}, {Suto}, {Takada}, {Takagi}, {Takata},
  {Takita}, {Tanaka}, {Terai}, {Toba}, {Uchiyama}, {Utsumi}, {Wang}, {Wang}, \&
  {Yamada}}]{2019PASJ...71..114A}
{Aihara}, H., {AlSayyad}, Y., {Ando}, M., {et~al.} 2019, \pasj, 71, 114

\bibitem[{{Ashby} {et~al.}(2018){Ashby}, {Caputi}, {Cowley}, {Deshmukh},
  {Dunlop}, {Milvang-Jensen}, {Fynbo}, {Muzzin}, {McCracken}, {Le F{\`e}vre},
  {Huang}, \& {Zhang}}]{2018ApJS..237...39A}
{Ashby}, M.~L.~N., {Caputi}, K.~I., {Cowley}, W., {et~al.} 2018, \apjs, 237, 39

\bibitem[{{Ba{\~n}ados} {et~al.}(2018){Ba{\~n}ados}, {Venemans},
  {Mazzucchelli}, {Farina}, {Walter}, {Wang}, {Decarli}, {Stern}, {Fan},
  {Davies}, {Hennawi}, {Simcoe}, {Turner}, {Rix}, {Yang}, {Kelson}, {Rudie}, \&
  {Winters}}]{2018Natur.553..473B}
{Ba{\~n}ados}, E., {Venemans}, B.~P., {Mazzucchelli}, C., {et~al.} 2018, \nat,
  553, 473

\bibitem[{{Behroozi} {et~al.}(2019){Behroozi}, {Wechsler}, {Hearin}, \&
  {Conroy}}]{2019MNRAS.488.3143B}
{Behroozi}, P., {Wechsler}, R.~H., {Hearin}, A.~P., \& {Conroy}, C. 2019,
  \mnras, 488, 3143

\bibitem[{{Behroozi} {et~al.}(2020){Behroozi}, {Conroy}, {Wechsler}, {Hearin},
  {Williams}, {Moster}, {Yung}, {Somerville}, {Gottl{\"o}ber}, {Yepes}, \&
  {Endsley}}]{2020MNRAS.499.5702B}
{Behroozi}, P., {Conroy}, C., {Wechsler}, R.~H., {et~al.} 2020, \mnras, 499,
  5702

\bibitem[{{Bertin} \& {Arnouts}(1996)}]{1996A&AS..117..393B}
{Bertin}, E., \& {Arnouts}, S. 1996, \aaps, 117, 393

\bibitem[{{Binggeli} {et~al.}(2021){Binggeli}, {Inoue}, {Hashimoto}, {Toribio},
  {Zackrisson}, {Ramstedt}, {Mawatari}, {Harikane}, {Matsuo}, {Okamoto}, {Ota},
  {Shimizu}, {Tamura}, {Taniguchi}, \& {Umehata}}]{2021A&A...646A..26B}
{Binggeli}, C., {Inoue}, A.~K., {Hashimoto}, T., {et~al.} 2021, \aap, 646, A26

\bibitem[{{Bouwens} {et~al.}(2020){Bouwens}, {Gonz{\'a}lez-L{\'o}pez},
  {Aravena}, {Decarli}, {Novak}, {Stefanon}, {Walter}, {Boogaard}, {Carilli},
  {Dudzevi{\v{c}}i{\={u}}t{\.{e}}}, {Smail}, {Daddi}, {da Cunha}, {Ivison},
  {Nanayakkara}, {Cortes}, {Cox}, {Inami}, {Oesch}, {Popping}, {Riechers}, {van
  der Werf}, {Weiss}, {Fudamoto}, \& {Wagg}}]{2020ApJ...902..112B}
{Bouwens}, R., {Gonz{\'a}lez-L{\'o}pez}, J., {Aravena}, M., {et~al.} 2020,
  \apj, 902, 112

\bibitem[{{Bouwens} {et~al.}(2015){Bouwens}, {Illingworth}, {Oesch}, {Trenti},
  {Labb{\'e}}, {Bradley}, {Carollo}, {van Dokkum}, {Gonzalez}, {Holwerda},
  {Franx}, {Spitler}, {Smit}, \& {Magee}}]{2015ApJ...803...34B}
{Bouwens}, R.~J., {Illingworth}, G.~D., {Oesch}, P.~A., {et~al.} 2015, \apj,
  803, 34

\bibitem[{{Bouwens} {et~al.}(2016){Bouwens}, {Oesch}, {Labb{\'e}},
  {Illingworth}, {Fazio}, {Coe}, {Holwerda}, {Smit}, {Stefanon}, {van Dokkum},
  {Trenti}, {Ashby}, {Huang}, {Spitler}, {Straatman}, {Bradley}, \&
  {Magee}}]{2016ApJ...830...67B}
{Bouwens}, R.~J., {Oesch}, P.~A., {Labb{\'e}}, I., {et~al.} 2016, \apj, 830, 67

\bibitem[{{Bouwens} {et~al.}(2021){Bouwens}, {Oesch}, {Stefanon},
  {Illingworth}, {Labb{\'e}}, {Reddy}, {Atek}, {Montes}, {Naidu},
  {Nanayakkara}, {Nelson}, \& {Wilkins}}]{2021AJ....162...47B}
{Bouwens}, R.~J., {Oesch}, P.~A., {Stefanon}, M., {et~al.} 2021, \aj, 162, 47

\bibitem[{{Bowler} {et~al.}(2020){Bowler}, {Jarvis}, {Dunlop}, {McLure},
  {McLeod}, {Adams}, {Milvang-Jensen}, \& {McCracken}}]{2020MNRAS.493.2059B}
{Bowler}, R.~A.~A., {Jarvis}, M.~J., {Dunlop}, J.~S., {et~al.} 2020, \mnras,
  493, 2059

\bibitem[{{Bruzual} \& {Charlot}(2003)}]{2003MNRAS.344.1000B}
{Bruzual}, G., \& {Charlot}, S. 2003, \mnras, 344, 1000

\bibitem[{{Calzetti} {et~al.}(2000){Calzetti}, {Armus}, {Bohlin}, {Kinney},
  {Koornneef}, \& {Storchi-Bergmann}}]{2000ApJ...533..682C}
{Calzetti}, D., {Armus}, L., {Bohlin}, R.~C., {et~al.} 2000, \apj, 533, 682

\bibitem[{{Carniani} {et~al.}(2017){Carniani}, {Maiolino}, {Pallottini},
  {Vallini}, {Pentericci}, {Ferrara}, {Castellano}, {Vanzella}, {Grazian},
  {Gallerani}, {Santini}, {Wagg}, \& {Fontana}}]{2017A&A...605A..42C}
{Carniani}, S., {Maiolino}, R., {Pallottini}, A., {et~al.} 2017, \aap, 605, A42

\bibitem[{{Chabrier}(2003)}]{2003PASP..115..763C}
{Chabrier}, G. 2003, \pasp, 115, 763

\bibitem[{{Chevallard} \& {Charlot}(2016)}]{2016MNRAS.462.1415C}
{Chevallard}, J., \& {Charlot}, S. 2016, \mnras, 462, 1415

\bibitem[{{Coleman} {et~al.}(1980){Coleman}, {Wu}, \&
  {Weedman}}]{1980ApJS...43..393C}
{Coleman}, G.~D., {Wu}, C.~C., \& {Weedman}, D.~W. 1980, \apjs, 43, 393

\bibitem[{{da Cunha} {et~al.}(2013){da Cunha}, {Groves}, {Walter}, {Decarli},
  {Weiss}, {Bertoldi}, {Carilli}, {Daddi}, {Elbaz}, {Ivison}, {Maiolino},
  {Riechers}, {Rix}, {Sargent}, \& {Smail}}]{2013ApJ...766...13D}
{da Cunha}, E., {Groves}, B., {Walter}, F., {et~al.} 2013, \apj, 766, 13

\bibitem[{{Davidzon} {et~al.}(2017){Davidzon}, {Ilbert}, {Laigle}, {Coupon},
  {McCracken}, {Delvecchio}, {Masters}, {Capak}, {Hsieh}, {Le F{\`e}vre},
  {Tresse}, {Bethermin}, {Chang}, {Faisst}, {Le Floc'h}, {Steinhardt}, {Toft},
  {Aussel}, {Dubois}, {Hasinger}, {Salvato}, {Sanders}, {Scoville}, \&
  {Silverman}}]{2017A&A...605A..70D}
{Davidzon}, I., {Ilbert}, O., {Laigle}, C., {et~al.} 2017, \aap, 605, A70

\bibitem[{{Dayal} {et~al.}(2013){Dayal}, {Dunlop}, {Maio}, \&
  {Ciardi}}]{2013MNRAS.434.1486D}
{Dayal}, P., {Dunlop}, J.~S., {Maio}, U., \& {Ciardi}, B. 2013, \mnras, 434,
  1486

\bibitem[{{Dayal} \& {Ferrara}(2018)}]{2018PhR...780....1D}
{Dayal}, P., \& {Ferrara}, A. 2018, \physrep, 780, 1

\bibitem[{{Dayal} {et~al.}(2014){Dayal}, {Ferrara}, {Dunlop}, \&
  {Pacucci}}]{2014MNRAS.445.2545D}
{Dayal}, P., {Ferrara}, A., {Dunlop}, J.~S., \& {Pacucci}, F. 2014, \mnras,
  445, 2545

\bibitem[{{Dayal} {et~al.}(2019){Dayal}, {Rossi}, {Shiralilou}, {Piana},
  {Choudhury}, \& {Volonteri}}]{2019MNRAS.486.2336D}
{Dayal}, P., {Rossi}, E.~M., {Shiralilou}, B., {et~al.} 2019, \mnras, 486, 2336

\bibitem[{{Dunlop} {et~al.}(2021){Dunlop}, {Abraham}, {Ashby}, {Bagley},
  {Best}, {Bongiorno}, {Bouwens}, {Bowler}, {Brammer}, {Bremer}, {Calabro'},
  {Carnall}, {Castellano}, {Cirasuolo}, {Conselice}, {Cullen}, {Dave}, {Dayal},
  {Dekel}, {Dickinson}, {Duncan}, {Elbaz}, {Ellis}, {Ferguson}, {Ferrara},
  {Finkelstein}, {Fontana}, {Furlanetto}, {Fynbo}, {Gallerani}, {Gardner},
  {Giavalisco}, {Grazian}, {Grogin}, {Harikane}, {Hopkins}, {Ilbert},
  {Illingworth}, {Juneau}, {Jung}, {Kartaltepe}, {Kassin}, {Kauffmann},
  {Khochfar}, {Kirkpatrick}, {Kocevski}, {Koekemoer}, {Labbe}, {Laporte},
  {Larson}, {Lucas}, {Magee}, {Mason}, {McCracken}, {McLeod}, {McLure},
  {Merlin}, {Mesinger}, {Milvang-Jensen}, {Newman}, {Oesch}, {Ouchi},
  {Pacifici}, {Papovich}, {Peacock}, {Peeples}, {Pentericci}, {Perez-Gonzalez},
  {Pirzkal}, {Pope}, {Pye}, {Reddy}, {Robertson}, {Salvato}, {Santini},
  {Schaerer}, {Shapley}, {Simons}, {Smit}, {Smith}, {Snyder}, {Somerville},
  {Stanway}, {Stefanon}, {Tasca}, {Tikkanen}, {Tresse}, {Trump}, {Whitaker},
  {Wilkins}, {Wright}, {Wyithe}, {van Dokkum}, \& {van der
  Werf}}]{2021jwst.prop.1837D}
{Dunlop}, J.~S., {Abraham}, R.~G., {Ashby}, M. L.~N., {et~al.} 2021, {PRIMER:
  Public Release IMaging for Extragalactic Research}, JWST Proposal. Cycle 1

\bibitem[{{Eisenstein} {et~al.}(2017{\natexlab{a}}){Eisenstein}, {Ferruit}, \&
  {Rieke}}]{2017jwst.prop.1181E}
{Eisenstein}, D.~J., {Ferruit}, P., \& {Rieke}, M.~J. 2017{\natexlab{a}},
  {NIRCam-NIRSpec galaxy assembly survey - GOODS-N}, JWST Proposal. Cycle 1

\bibitem[{{Eisenstein} {et~al.}(2017{\natexlab{b}}){Eisenstein}, {Ferruit},
  {Rieke}, {Willmer}, \& {Willott}}]{2017jwst.prop.1180E}
{Eisenstein}, D.~J., {Ferruit}, P., {Rieke}, M.~J., {Willmer}, C. N.~A., \&
  {Willott}, C.~J. 2017{\natexlab{b}}, {NIRCam-NIRSpec galaxy assembly survey -
  GOODS-S - part \#1a}, JWST Proposal. Cycle 1

\bibitem[{{Ellis} {et~al.}(2013){Ellis}, {McLure}, {Dunlop}, {Robertson},
  {Ono}, {Schenker}, {Koekemoer}, {Bowler}, {Ouchi}, {Rogers}, {Curtis-Lake},
  {Schneider}, {Charlot}, {Stark}, {Furlanetto}, \&
  {Cirasuolo}}]{2013ApJ...763L...7E}
{Ellis}, R.~S., {McLure}, R.~J., {Dunlop}, J.~S., {et~al.} 2013, \apjl, 763, L7

\bibitem[{{Finkelstein} {et~al.}(2015){Finkelstein}, {Ryan}, {Papovich},
  {Dickinson}, {Song}, {Somerville}, {Ferguson}, {Salmon}, {Giavalisco},
  {Koekemoer}, {Ashby}, {Behroozi}, {Castellano}, {Dunlop}, {Faber}, {Fazio},
  {Fontana}, {Grogin}, {Hathi}, {Jaacks}, {Kocevski}, {Livermore}, {McLure},
  {Merlin}, {Mobasher}, {Newman}, {Rafelski}, {Tilvi}, \&
  {Willner}}]{2015ApJ...810...71F}
{Finkelstein}, S.~L., {Ryan}, Jr., R.~E., {Papovich}, C., {et~al.} 2015, \apj,
  810, 71

\bibitem[{{Finkelstein} {et~al.}(2017){Finkelstein}, {Dickinson}, {Ferguson},
  {Grazian}, {Grogin}, {Kartaltepe}, {Kewley}, {Kocevski}, {Koekemoer}, {Lotz},
  {Papovich}, {Pentericci}, {Perez-Gonzalez}, {Pirzkal}, {Ravindranath},
  {Somerville}, {Trump}, \& {Wilkins}}]{2017jwst.prop.1345F}
{Finkelstein}, S.~L., {Dickinson}, M., {Ferguson}, H.~C., {et~al.} 2017, {The
  Cosmic Evolution Early Release Science (CEERS) Survey}, JWST Proposal ID
  1345. Cycle 0 Early Release Science

\bibitem[{{Finkelstein} {et~al.}(2021{\natexlab{a}}){Finkelstein}, {Bagley},
  {Song}, {Larson}, {Papovich}, {Dickinson}, {Finkelstein}, {Koekemoer},
  {Pirzkal}, {Somerville}, {Yung}, {Behroozi}, {Ferguson}, {Giavalisco},
  {Grogin}, {Hathi}, {Hutchison}, {Jung}, {Kocevski}, {Kawinwanichakij},
  {Rojas-Ruiz}, {Ryan}, {Snyder}, \& {Tacchella}}]{2021arXiv210613813F}
{Finkelstein}, S.~L., {Bagley}, M., {Song}, M., {et~al.} 2021{\natexlab{a}},
  arXiv e-prints, arXiv:2106.13813

\bibitem[{{Finkelstein} {et~al.}(2021{\natexlab{b}}){Finkelstein}, {Papovich},
  {Pirzkal}, {Bagley}, {Berg}, {Castellano}, {Chavez Ortiz}, {Chworowsky},
  {Dave}, {Dickinson}, {Estrada-Carpenter}, {Faber}, {Ferguson}, {Fontana},
  {Giavalisco}, {Grazian}, {Grogin}, {Jaskot}, {Jung}, {Kartaltepe}, {Kewley},
  {Kirkpatrick}, {Kocevski}, {Larson}, {Leung}, {Lotz}, {Mantha}, {Matharu},
  {McCarron}, {McIntosh}, {Natarajan}, {Pentericci}, {Ravindranath},
  {Rodriguez-Gomez}, {Rothberg}, {Ryan}, {Simons}, {Snyder}, {Somerville},
  {Trump}, {Wilkins}, \& {Yung}}]{2021jwst.prop.2079F}
{Finkelstein}, S.~L., {Papovich}, C., {Pirzkal}, N., {et~al.}
  2021{\natexlab{b}}, {The Webb Deep Extragalactic Exploratory Public (WDEEP)
  Survey: Feedback in Low-Mass Galaxies from Cosmic Dawn to Dusk}, JWST
  Proposal. Cycle 1

\bibitem[{{Fudamoto} {et~al.}(2020){Fudamoto}, {Oesch}, {Faisst},
  {B{\'e}thermin}, {Ginolfi}, {Khusanova}, {Loiacono}, {Le F{\`e}vre}, {Capak},
  {Schaerer}, {Silverman}, {Cassata}, {Yan}, {Amorin}, {Bardelli}, {Boquien},
  {Cimatti}, {Dessauges-Zavadsky}, {Fujimoto}, {Gruppioni}, {Hathi}, {Ibar},
  {Jones}, {Koekemoer}, {Lagache}, {Lemaux}, {Maiolino}, {Narayanan}, {Pozzi},
  {Riechers}, {Rodighiero}, {Talia}, {Toft}, {Vallini}, {Vergani}, {Zamorani},
  \& {Zucca}}]{2020A&A...643A...4F}
{Fudamoto}, Y., {Oesch}, P.~A., {Faisst}, A., {et~al.} 2020, \aap, 643, A4

\bibitem[{{Furusawa} {et~al.}(2008){Furusawa}, {Kosugi}, {Akiyama}, {Takata},
  {Sekiguchi}, {Tanaka}, {Iwata}, {Kajisawa}, {Yasuda}, {Doi}, {Ouchi},
  {Simpson}, {Shimasaku}, {Yamada}, {Furusawa}, {Morokuma}, {Ishida}, {Aoki},
  {Fuse}, {Imanishi}, {Iye}, {Karoji}, {Kobayashi}, {Kodama}, {Komiyama},
  {Maeda}, {Miyazaki}, {Mizumoto}, {Nakata}, {Noumaru}, {Ogasawara}, {Okamura},
  {Saito}, {Sasaki}, {Ueda}, \& {Yoshida}}]{2008ApJS..176....1F}
{Furusawa}, H., {Kosugi}, G., {Akiyama}, M., {et~al.} 2008, \apjs, 176, 1

\bibitem[{{Gehrels}(1986)}]{1986ApJ...303..336G}
{Gehrels}, N. 1986, \apj, 303, 336

\bibitem[{{Glazebrook} {et~al.}(2017){Glazebrook}, {Schreiber}, {Labb{\'e}},
  {Nanayakkara}, {Kacprzak}, {Oesch}, {Papovich}, {Spitler}, {Straatman},
  {Tran}, \& {Yuan}}]{2017Natur.544...71G}
{Glazebrook}, K., {Schreiber}, C., {Labb{\'e}}, I., {et~al.} 2017, \nat, 544,
  71

\bibitem[{{Haiman} \& {Menou}(2000)}]{2000ApJ...531...42H}
{Haiman}, Z., \& {Menou}, K. 2000, \apj, 531, 42

\bibitem[{{Harikane} {et~al.}(2021{\natexlab{a}}){Harikane}, {Fudamoto},
  {Hashimoto}, {Inoue}, {Matsuo}, {Tamura}, \&
  {Yamanaka}}]{2021jwst.prop.1740H}
{Harikane}, Y., {Fudamoto}, Y., {Hashimoto}, T., {et~al.} 2021{\natexlab{a}},
  {H-drop galaxies: ``Rosetta Stones'' at z 13 for galaxy formation studies},
  JWST Proposal. Cycle 1

\bibitem[{{Harikane} {et~al.}(2018){Harikane}, {Ouchi}, {Shibuya}, {Kojima},
  {Zhang}, {Itoh}, {Ono}, {Higuchi}, {Inoue}, {Chevallard}, {Capak}, {Nagao},
  {Onodera}, {Faisst}, {Martin}, {Rauch}, {Bruzual}, {Charlot}, {Davidzon},
  {Fujimoto}, {Hilmi}, {Ilbert}, {Lee}, {Matsuoka}, {Silverman}, \&
  {Toft}}]{2018ApJ...859...84H}
{Harikane}, Y., {Ouchi}, M., {Shibuya}, T., {et~al.} 2018, \apj, 859, 84

\bibitem[{{Harikane} {et~al.}(2019){Harikane}, {Ouchi}, {Ono}, {Fujimoto},
  {Donevski}, {Shibuya}, {Faisst}, {Goto}, {Hatsukade}, {Kashikawa}, {Kohno},
  {Hashimoto}, {Higuchi}, {Inoue}, {Lin}, {Martin}, {Overzier}, {Smail},
  {Toshikawa}, {Umehata}, {Ao}, {Chapman}, {Clements}, {Im}, {Jing},
  {Kawaguchi}, {Lee}, {Lee}, {Lin}, {Matsuoka}, {Marinello}, {Nagao},
  {Onodera}, {Toft}, \& {Wang}}]{2019ApJ...883..142H}
{Harikane}, Y., {Ouchi}, M., {Ono}, Y., {et~al.} 2019, \apj, 883, 142

\bibitem[{{Harikane} {et~al.}(2020){Harikane}, {Ouchi}, {Inoue}, {Matsuoka},
  {Tamura}, {Bakx}, {Fujimoto}, {Moriwaki}, {Ono}, {Nagao}, {Tadaki}, {Kojima},
  {Shibuya}, {Egami}, {Ferrara}, {Gallerani}, {Hashimoto}, {Kohno}, {Matsuda},
  {Matsuo}, {Pallottini}, {Sugahara}, \& {Vallini}}]{2020ApJ...896...93H}
{Harikane}, Y., {Ouchi}, M., {Inoue}, A.~K., {et~al.} 2020, \apj, 896, 93

\bibitem[{{Harikane} {et~al.}(2021{\natexlab{b}}){Harikane}, {Ono}, {Ouchi},
  {Liu}, {Sawicki}, {Shibuya}, {Behroozi}, {He}, {Shimasaku}, {Arnouts},
  {Coupon}, {Fujimoto}, {Gwyn}, {Huang}, {Inoue}, {Kashikawa}, {Komiyama},
  {Matsuoka}, \& {Willott}}]{2021arXiv210801090H}
{Harikane}, Y., {Ono}, Y., {Ouchi}, M., {et~al.} 2021{\natexlab{b}}, arXiv
  e-prints, arXiv:2108.01090

\bibitem[{{Hashimoto} {et~al.}(2018){Hashimoto}, {Laporte}, {Mawatari},
  {Ellis}, {Inoue}, {Zackrisson}, {Roberts-Borsani}, {Zheng}, {Tamura},
  {Bauer}, {Fletcher}, {Harikane}, {Hatsukade}, {Hayatsu}, {Matsuda}, {Matsuo},
  {Okamoto}, {Ouchi}, {Pell{\'o}}, {Rydberg}, {Shimizu}, {Taniguchi},
  {Umehata}, \& {Yoshida}}]{2018Natur.557..392H}
{Hashimoto}, T., {Laporte}, N., {Mawatari}, K., {et~al.} 2018, \nat, 557, 392

\bibitem[{{Hashimoto} {et~al.}(2019){Hashimoto}, {Inoue}, {Mawatari}, {Tamura},
  {Matsuo}, {Furusawa}, {Harikane}, {Shibuya}, {Knudsen}, {Kohno}, {Ono},
  {Zackrisson}, {Okamoto}, {Kashikawa}, {Oesch}, {Ouchi}, {Ota}, {Shimizu},
  {Taniguchi}, {Umehata}, \& {Watson}}]{2019PASJ...71...71H}
{Hashimoto}, T., {Inoue}, A.~K., {Mawatari}, K., {et~al.} 2019, \pasj, 71, 71

\bibitem[{{Hutter} {et~al.}(2021){Hutter}, {Dayal}, {Yepes}, {Gottl{\"o}ber},
  {Legrand}, \& {Ucci}}]{2021MNRAS.503.3698H}
{Hutter}, A., {Dayal}, P., {Yepes}, G., {et~al.} 2021, \mnras, 503, 3698

\bibitem[{{Inoue}(2011)}]{2011MNRAS.415.2920I}
{Inoue}, A.~K. 2011, \mnras, 415, 2920

\bibitem[{{Inoue} {et~al.}(2014{\natexlab{a}}){Inoue}, {Shimizu}, {Iwata}, \&
  {Tanaka}}]{2014MNRAS.442.1805I}
{Inoue}, A.~K., {Shimizu}, I., {Iwata}, I., \& {Tanaka}, M. 2014{\natexlab{a}},
  \mnras, 442, 1805

\bibitem[{{Inoue} {et~al.}(2014{\natexlab{b}}){Inoue}, {Shimizu}, {Tamura},
  {Matsuo}, {Okamoto}, \& {Yoshida}}]{2014ApJ...780L..18I}
{Inoue}, A.~K., {Shimizu}, I., {Tamura}, Y., {et~al.} 2014{\natexlab{b}},
  \apjl, 780, L18

\bibitem[{{Inoue} {et~al.}(2016){Inoue}, {Tamura}, {Matsuo}, {Mawatari},
  {Shimizu}, {Shibuya}, {Ota}, {Yoshida}, {Zackrisson}, {Kashikawa}, {Kohno},
  {Umehata}, {Hatsukade}, {Iye}, {Matsuda}, {Okamoto}, \&
  {Yamaguchi}}]{2016Sci...352.1559I}
{Inoue}, A.~K., {Tamura}, Y., {Matsuo}, H., {et~al.} 2016, Science, 352, 1559

\bibitem[{{Jansen} \& {Windhorst}(2018)}]{2018PASP..130l4001J}
{Jansen}, R.~A., \& {Windhorst}, R.~A. 2018, \pasp, 130, 124001

\bibitem[{{Jarvis} {et~al.}(2013){Jarvis}, {Bonfield}, {Bruce}, {Geach},
  {McAlpine}, {McLure}, {Gonz{\'a}lez-Solares}, {Irwin}, {Lewis}, {Yoldas},
  {Andreon}, {Cross}, {Emerson}, {Dalton}, {Dunlop}, {Hodgkin}, {Le},
  {Karouzos}, {Meisenheimer}, {Oliver}, {Rawlings}, {Simpson}, {Smail},
  {Smith}, {Sullivan}, {Sutherland}, {White}, \& {Zwart}}]{2013MNRAS.428.1281J}
{Jarvis}, M.~J., {Bonfield}, D.~G., {Bruce}, V.~A., {et~al.} 2013, \mnras, 428,
  1281

\bibitem[{{Jiang} {et~al.}(2021){Jiang}, {Kashikawa}, {Wang}, {Walth}, {Ho},
  {Cai}, {Egami}, {Fan}, {Ito}, {Liang}, {Schaerer}, \&
  {Stark}}]{2021NatAs...5..256J}
{Jiang}, L., {Kashikawa}, N., {Wang}, S., {et~al.} 2021, Nature Astronomy, 5,
  256

\bibitem[{{Kartaltepe} {et~al.}(2021){Kartaltepe}, {Casey}, {Bagley},
  {Bongiorno}, {Capak}, {Champagne}, {Cooke}, {Cooper}, {Darvish}, {Davidzon},
  {Drakos}, {Drew}, {Faisst}, {Finkelstein}, {Hayward}, {Hemmati},
  {Hirschmann}, {Ilbert}, {Jahnke}, {Koekemoer}, {Liu}, {Long}, {Magdis},
  {Manning}, {Maraston}, {Martin}, {Massey}, {McCleary}, {McCracken},
  {Nayyeri}, {Renzini}, {Rhodes}, {Rich}, {Robertson}, {Rose}, {Sanders},
  {Scarlata}, {Scoville}, {Sheth}, {Silverman}, {Sparre}, {Talia}, {Toft},
  {Trakhtenbrot}, {Vanderhoof}, {Vardoulaki}, {Whitaker}, {Wilkins}, \&
  {Zavala}}]{2021jwst.prop.1727K}
{Kartaltepe}, J., {Casey}, C.~M., {Bagley}, M., {et~al.} 2021, {COSMOS-Webb:
  The Webb Cosmic Origins Survey}, JWST Proposal. Cycle 1

\bibitem[{{Kirkpatrick} {et~al.}(2011){Kirkpatrick}, {Cushing}, {Gelino},
  {Griffith}, {Skrutskie}, {Marsh}, {Wright}, {Mainzer}, {Eisenhardt},
  {McLean}, {Thompson}, {Bauer}, {Benford}, {Bridge}, {Lake}, {Petty},
  {Stanford}, {Tsai}, {Bailey}, {Beichman}, {Bloom}, {Bochanski}, {Burgasser},
  {Capak}, {Cruz}, {Hinz}, {Kartaltepe}, {Knox}, {Manohar}, {Masters},
  {Morales-Calder{\'o}n}, {Prato}, {Rodigas}, {Salvato}, {Schurr}, {Scoville},
  {Simcoe}, {Stapelfeldt}, {Stern}, {Stock}, \& {Vacca}}]{2011ApJS..197...19K}
{Kirkpatrick}, J.~D., {Cushing}, M.~C., {Gelino}, C.~R., {et~al.} 2011, \apjs,
  197, 19

\bibitem[{{Koekemoer} {et~al.}(2019){Koekemoer}, {Foley}, {Spergel}, {Bagley},
  {Bezanson}, {Bianco}, {Capak}, {De Rosa}, {Dickinson}, {Dore}, {Fan},
  {Fazio}, {Ferguson}, {Filippenko}, {Finkelstein}, {Frye}, {Gawiser},
  {Grogin}, {Hathi}, {Hirata}, {Hounsell}, {Jansen}, {Jha}, {Kartaltepe},
  {Kim}, {Kelly}, {Kruk}, {Larson}, {Lucas}, {Malhotra}, {Mandel}, {Margutti},
  {Marrone}, {McQuinn}, {Melchior}, {Moustakas}, {Newman}, {Papovich},
  {Peeples}, {Perlmutter}, {Rhoads}, {Rhodes}, {Robertson}, {Rubin}, {Ryan},
  {Scolnic}, {Shapley}, {Somerville}, {Street}, {Wang}, {Whalen}, {Windhorst},
  \& {Wollack}}]{2019BAAS...51c.550K}
{Koekemoer}, A., {Foley}, R.~J., {Spergel}, D.~N., {et~al.} 2019, \baas, 51,
  550

\bibitem[{{Labbe} {et~al.}(2021){Labbe}, {Bezanson}, {Atek}, {Brammer}, {Coe},
  {Dayal}, {Feldmann}, {Forster Schreiber}, {Franx}, {Geha}, {Glazebrook},
  {Greene}, {Juneau}, {Kassin}, {Kriek}, {Leja}, {Marchesini}, {Maseda},
  {Mowla}, {Muzzin}, {Nanayakkara}, {Nelson}, {Oesch}, {Pacifici}, {Papovich},
  {Price}, {Shapley}, {Stefanon}, {Taylor}, {Whitaker}, {Williams}, {Zitrin},
  {de Graaff}, \& {van Dokkum}}]{2021jwst.prop.2561L}
{Labbe}, I., {Bezanson}, R., {Atek}, H., {et~al.} 2021, {UNCOVER: Ultra-deep
  NIRCam and NIRSpec Observations Before the Epoch of Reionization}, JWST
  Proposal. Cycle 1

\bibitem[{{Laigle} {et~al.}(2016){Laigle}, {McCracken}, {Ilbert}, {Hsieh},
  {Davidzon}, {Capak}, {Hasinger}, {Silverman}, {Pichon}, {Coupon}, {Aussel},
  {Le Borgne}, {Caputi}, {Cassata}, {Chang}, {Civano}, {Dunlop}, {Fynbo},
  {Kartaltepe}, {Koekemoer}, {Le F{\`e}vre}, {Le Floc'h}, {Leauthaud}, {Lilly},
  {Lin}, {Marchesi}, {Milvang-Jensen}, {Salvato}, {Sanders}, {Scoville},
  {Smolcic}, {Stockmann}, {Taniguchi}, {Tasca}, {Toft}, {Vaccari}, \&
  {Zabl}}]{2016ApJS..224...24L}
{Laigle}, C., {McCracken}, H.~J., {Ilbert}, O., {et~al.} 2016, \apjs, 224, 24

\bibitem[{{Laporte} {et~al.}(2021){Laporte}, {Meyer}, {Ellis}, {Robertson},
  {Chisholm}, \& {Roberts-Borsani}}]{2021MNRAS.505.3336L}
{Laporte}, N., {Meyer}, R.~A., {Ellis}, R.~S., {et~al.} 2021, \mnras, 505, 3336

\bibitem[{{Laporte} {et~al.}(2017){Laporte}, {Ellis}, {Boone}, {Bauer},
  {Qu{\'e}nard}, {Roberts-Borsani}, {Pell{\'o}}, {P{\'e}rez-Fournon}, \&
  {Streblyanska}}]{2017ApJ...837L..21L}
{Laporte}, N., {Ellis}, R.~S., {Boone}, F., {et~al.} 2017, \apjl, 837, L21

\bibitem[{{Lawrence} {et~al.}(2007){Lawrence}, {Warren}, {Almaini}, {Edge},
  {Hambly}, {Jameson}, {Lucas}, {Casali}, {Adamson}, {Dye}, {Emerson},
  {Foucaud}, {Hewett}, {Hirst}, {Hodgkin}, {Irwin}, {Lodieu}, {McMahon},
  {Simpson}, {Smail}, {Mortlock}, \& {Folger}}]{2007MNRAS.379.1599L}
{Lawrence}, A., {Warren}, S.~J., {Almaini}, O., {et~al.} 2007, \mnras, 379,
  1599

\bibitem[{{Madau} \& {Dickinson}(2014)}]{2014ARA&A..52..415M}
{Madau}, P., \& {Dickinson}, M. 2014, \araa, 52, 415

\bibitem[{{Marrone} {et~al.}(2018){Marrone}, {Spilker}, {Hayward}, {Vieira},
  {Aravena}, {Ashby}, {Bayliss}, {B{\'e}thermin}, {Brodwin}, {Bothwell},
  {Carlstrom}, {Chapman}, {Chen}, {Crawford}, {Cunningham}, {De Breuck},
  {Fassnacht}, {Gonzalez}, {Greve}, {Hezaveh}, {Lacaille}, {Litke}, {Lower},
  {Ma}, {Malkan}, {Miller}, {Morningstar}, {Murphy}, {Narayanan}, {Phadke},
  {Rotermund}, {Sreevani}, {Stalder}, {Stark}, {Strand et}, {Tang}, \&
  {Wei{\ss}}}]{2018Natur.553...51M}
{Marrone}, D.~P., {Spilker}, J.~S., {Hayward}, C.~C., {et~al.} 2018, \nat, 553,
  51

\bibitem[{{Mason} {et~al.}(2015){Mason}, {Trenti}, \&
  {Treu}}]{2015ApJ...813...21M}
{Mason}, C.~A., {Trenti}, M., \& {Treu}, T. 2015, \apj, 813, 21

\bibitem[{{Mawatari} {et~al.}(2020{\natexlab{a}}){Mawatari}, {Inoue},
  {Yamanaka}, {Hashimoto}, \& {Tamura}}]{2020IAUS..341..285M}
{Mawatari}, K., {Inoue}, A.~K., {Yamanaka}, S., {Hashimoto}, T., \& {Tamura},
  Y. 2020{\natexlab{a}}, in Panchromatic Modelling with Next Generation
  Facilities, ed. M.~{Boquien}, E.~{Lusso}, C.~{Gruppioni}, \& P.~{Tissera},
  Vol. 341, 285--286

\bibitem[{{Mawatari} {et~al.}(2020{\natexlab{b}}){Mawatari}, {Inoue},
  {Hashimoto}, {Silverman}, {Kajisawa}, {Yamanaka}, {Yamada}, {Davidzon},
  {Capak}, {Lin}, {Hsieh}, {Taniguchi}, {Tanaka}, {Ono}, {Harikane},
  {Sugahara}, {Fujimoto}, \& {Nagao}}]{2020ApJ...889..137M}
{Mawatari}, K., {Inoue}, A.~K., {Hashimoto}, T., {et~al.} 2020{\natexlab{b}},
  \apj, 889, 137

\bibitem[{{McCracken} {et~al.}(2012){McCracken}, {Milvang-Jensen}, {Dunlop},
  {Franx}, {Fynbo}, {Le F{\`e}vre}, {Holt}, {Caputi}, {Goranova}, {Buitrago},
  {Emerson}, {Freudling}, {Hudelot}, {L{\'o}pez-Sanjuan}, {Magnard}, {Mellier},
  {M{\o}ller}, {Nilsson}, {Sutherland}, {Tasca}, \&
  {Zabl}}]{2012A&A...544A.156M}
{McCracken}, H.~J., {Milvang-Jensen}, B., {Dunlop}, J., {et~al.} 2012, \aap,
  544, A156

\bibitem[{{McLeod} {et~al.}(2016){McLeod}, {McLure}, \&
  {Dunlop}}]{2016MNRAS.459.3812M}
{McLeod}, D.~J., {McLure}, R.~J., \& {Dunlop}, J.~S. 2016, \mnras, 459, 3812

\bibitem[{{McMullin} {et~al.}(2007){McMullin}, {Waters}, {Schiebel}, {Young},
  \& {Golap}}]{2007ASPC..376..127M}
{McMullin}, J.~P., {Waters}, B., {Schiebel}, D., {Young}, W., \& {Golap}, K.
  2007, in Astronomical Society of the Pacific Conference Series, Vol. 376,
  Astronomical Data Analysis Software and Systems XVI, ed. R.~A. {Shaw},
  F.~{Hill}, \& D.~J. {Bell}, 127

\bibitem[{{Mehta} {et~al.}(2018){Mehta}, {Scarlata}, {Capak}, {Davidzon},
  {Faisst}, {Hsieh}, {Ilbert}, {Jarvis}, {Laigle}, {Phillips}, {Silverman},
  {Strauss}, {Tanaka}, {Bowler}, {Coupon}, {Foucaud}, {Hemmati}, {Masters},
  {McCracken}, {Mobasher}, {Ouchi}, {Shibuya}, \& {Wang}}]{2018ApJS..235...36M}
{Mehta}, V., {Scarlata}, C., {Capak}, P., {et~al.} 2018, \apjs, 235, 36

\bibitem[{{Merlin} {et~al.}(2016){Merlin}, {Bourne}, {Castellano}, {Ferguson},
  {Wang}, {Derriere}, {Dunlop}, {Elbaz}, \& {Fontana}}]{2016A&A...595A..97M}
{Merlin}, E., {Bourne}, N., {Castellano}, M., {et~al.} 2016, \aap, 595, A97

\bibitem[{{Morishita} {et~al.}(2018){Morishita}, {Trenti}, {Stiavelli},
  {Bradley}, {Coe}, {Oesch}, {Mason}, {Bridge}, {Holwerda}, {Livermore},
  {Salmon}, {Schmidt}, {Shull}, \& {Treu}}]{2018ApJ...867..150M}
{Morishita}, T., {Trenti}, M., {Stiavelli}, M., {et~al.} 2018, \apj, 867, 150

\bibitem[{{Mortlock} {et~al.}(2011){Mortlock}, {Warren}, {Venemans}, {Patel},
  {Hewett}, {McMahon}, {Simpson}, {Theuns}, {Gonz{\'a}les-Solares}, {Adamson},
  {Dye}, {Hambly}, {Hirst}, {Irwin}, {Kuiper}, {Lawrence}, \&
  {R{\"o}ttgering}}]{2011Natur.474..616M}
{Mortlock}, D.~J., {Warren}, S.~J., {Venemans}, B.~P., {et~al.} 2011, \nat,
  474, 616

\bibitem[{{Oesch} {et~al.}(2018){Oesch}, {Bouwens}, {Illingworth}, {Labb{\'e}},
  \& {Stefanon}}]{2018ApJ...855..105O}
{Oesch}, P.~A., {Bouwens}, R.~J., {Illingworth}, G.~D., {Labb{\'e}}, I., \&
  {Stefanon}, M. 2018, \apj, 855, 105

\bibitem[{{Oesch} {et~al.}(2016){Oesch}, {Brammer}, {van Dokkum},
  {Illingworth}, {Bouwens}, {Labb{\'e}}, {Franx}, {Momcheva}, {Ashby}, {Fazio},
  {Gonzalez}, {Holden}, {Magee}, {Skelton}, {Smit}, {Spitler}, {Trenti}, \&
  {Willner}}]{2016ApJ...819..129O}
{Oesch}, P.~A., {Brammer}, G., {van Dokkum}, P.~G., {et~al.} 2016, \apj, 819,
  129

\bibitem[{{Oke} \& {Gunn}(1983)}]{1983ApJ...266..713O}
{Oke}, J.~B., \& {Gunn}, J.~E. 1983, \apj, 266, 713

\bibitem[{{Oliver} {et~al.}(2012){Oliver}, {Bock}, {Altieri}, {Amblard},
  {Arumugam}, {Aussel}, {Babbedge}, {Beelen}, {B{\'e}thermin}, {Blain},
  {Boselli}, {Bridge}, {Brisbin}, {Buat}, {Burgarella},
  {Castro-Rodr{\'\i}guez}, {Cava}, {Chanial}, {Cirasuolo}, {Clements},
  {Conley}, {Conversi}, {Cooray}, {Dowell}, {Dubois}, {Dwek}, {Dye}, {Eales},
  {Elbaz}, {Farrah}, {Feltre}, {Ferrero}, {Fiolet}, {Fox}, {Franceschini},
  {Gear}, {Giovannoli}, {Glenn}, {Gong}, {Gonz{\'a}lez Solares}, {Griffin},
  {Halpern}, {Harwit}, {Hatziminaoglou}, {Heinis}, {Hurley}, {Hwang}, {Hyde},
  {Ibar}, {Ilbert}, {Isaak}, {Ivison}, {Lagache}, {Le Floc'h}, {Levenson},
  {Faro}, {Lu}, {Madden}, {Maffei}, {Magdis}, {Mainetti}, {Marchetti},
  {Marsden}, {Marshall}, {Mortier}, {Nguyen}, {O'Halloran}, {Omont}, {Page},
  {Panuzzo}, {Papageorgiou}, {Patel}, {Pearson}, {P{\'e}rez-Fournon}, {Pohlen},
  {Rawlings}, {Raymond}, {Rigopoulou}, {Riguccini}, {Rizzo}, {Rodighiero},
  {Roseboom}, {Rowan-Robinson}, {S{\'a}nchez Portal}, {Schulz}, {Scott},
  {Seymour}, {Shupe}, {Smith}, {Stevens}, {Symeonidis}, {Trichas}, {Tugwell},
  {Vaccari}, {Valtchanov}, {Vieira}, {Viero}, {Vigroux}, {Wang}, {Ward},
  {Wardlow}, {Wright}, {Xu}, \& {Zemcov}}]{2012MNRAS.424.1614O}
{Oliver}, S.~J., {Bock}, J., {Altieri}, B., {et~al.} 2012, \mnras, 424, 1614

\bibitem[{{Ono} {et~al.}(2018){Ono}, {Ouchi}, {Harikane}, {Toshikawa}, {Rauch},
  {Yuma}, {Sawicki}, {Shibuya}, {Shimasaku}, {Oguri}, {Willott}, {Akhlaghi},
  {Akiyama}, {Coupon}, {Kashikawa}, {Komiyama}, {Konno}, {Lin}, {Matsuoka},
  {Miyazaki}, {Nagao}, {Nakajima}, {Silverman}, {Tanaka}, {Taniguchi}, \&
  {Wang}}]{2018PASJ...70S..10O}
{Ono}, Y., {Ouchi}, M., {Harikane}, Y., {et~al.} 2018, \pasj, 70, S10

\bibitem[{{Pacucci} {et~al.}(2022){Pacucci}, {Dayal}, {Harikane}, {Inoue}, \&
  {Loeb}}]{2022arXiv220100823P}
{Pacucci}, F., {Dayal}, P., {Harikane}, Y., {Inoue}, A.~K., \& {Loeb}, A. 2022,
  arXiv e-prints, arXiv:2201.00823

\bibitem[{{Patten} {et~al.}(2006){Patten}, {Stauffer}, {Burrows}, {Marengo},
  {Hora}, {Luhman}, {Sonnett}, {Henry}, {Raghavan}, {Megeath}, {Liebert}, \&
  {Fazio}}]{2006ApJ...651..502P}
{Patten}, B.~M., {Stauffer}, J.~R., {Burrows}, A., {et~al.} 2006, \apj, 651,
  502

\bibitem[{{Piana} {et~al.}(2021){Piana}, {Dayal}, \&
  {Choudhury}}]{2021arXiv211103105P}
{Piana}, O., {Dayal}, P., \& {Choudhury}, T.~R. 2021, arXiv e-prints,
  arXiv:2111.03105

\bibitem[{{Planck Collaboration} {et~al.}(2016){Planck Collaboration}, {Ade},
  {Aghanim}, {Arnaud}, {Ashdown}, {Aumont}, {Baccigalupi}, {Banday},
  {Barreiro}, {Bartlett}, \& et~al.}]{2016A&A...594A..13P}
{Planck Collaboration}, {Ade}, P.~A.~R., {Aghanim}, N., {et~al.} 2016, \aap,
  594, A13

\bibitem[{{Rhoads} {et~al.}(2018){Rhoads}, {Malhotra}, {Jansen}, {Windhorst},
  {Tilvi}, {Finkelstein}, {Wold}, {Papovich}, {Fan}, {Mellema}, {Zackrisson},
  {Jensen}, \& {T}}]{2018AAS...23125817R}
{Rhoads}, J., {Malhotra}, S., {Jansen}, R.~A., {et~al.} 2018, in American
  Astronomical Society Meeting Abstracts, Vol. 231, American Astronomical
  Society Meeting Abstracts \#231, 258.17

\bibitem[{{Rieke} {et~al.}(2005){Rieke}, {Kelly}, \&
  {Horner}}]{2005SPIE.5904....1R}
{Rieke}, M.~J., {Kelly}, D., \& {Horner}, S. 2005, in Society of Photo-Optical
  Instrumentation Engineers (SPIE) Conference Series, Vol. 5904, Cryogenic
  Optical Systems and Instruments XI, ed. J.~B. {Heaney} \& L.~G. {Burriesci},
  1--8

\bibitem[{{Roberts-Borsani} {et~al.}(2021{\natexlab{a}}){Roberts-Borsani},
  {Morishita}, {Treu}, {Leethochawalit}, \& {Trenti}}]{2021arXiv210606544R}
{Roberts-Borsani}, G., {Morishita}, T., {Treu}, T., {Leethochawalit}, N., \&
  {Trenti}, M. 2021{\natexlab{a}}, arXiv e-prints, arXiv:2106.06544

\bibitem[{{Roberts-Borsani} {et~al.}(2021{\natexlab{b}}){Roberts-Borsani},
  {Treu}, {Mason}, {Schmidt}, {Jones}, \& {Fontana}}]{2021ApJ...910...86R}
{Roberts-Borsani}, G., {Treu}, T., {Mason}, C., {et~al.} 2021{\natexlab{b}},
  \apj, 910, 86

\bibitem[{{Roberts-Borsani} {et~al.}(2020){Roberts-Borsani}, {Ellis}, \&
  {Laporte}}]{2020MNRAS.497.3440R}
{Roberts-Borsani}, G.~W., {Ellis}, R.~S., \& {Laporte}, N. 2020, \mnras, 497,
  3440

\bibitem[{{Robertson}(2021)}]{2021arXiv211013160R}
{Robertson}, B.~E. 2021, arXiv e-prints, arXiv:2110.13160

\bibitem[{{Rogers} {et~al.}(2014){Rogers}, {McLure}, {Dunlop}, {Bowler},
  {Curtis-Lake}, {Dayal}, {Faber}, {Ferguson}, {Finkelstein}, {Grogin},
  {Hathi}, {Kocevski}, {Koekemoer}, \& {Kurczynski}}]{2014MNRAS.440.3714R}
{Rogers}, A.~B., {McLure}, R.~J., {Dunlop}, J.~S., {et~al.} 2014, \mnras, 440,
  3714

\bibitem[{{Salpeter}(1955)}]{1955ApJ...121..161S}
{Salpeter}, E.~E. 1955, \apj, 121, 161

\bibitem[{{Sawicki}(2012)}]{2012PASP..124.1208S}
{Sawicki}, M. 2012, \pasp, 124, 1208

\bibitem[{{Scoville} {et~al.}(2007){Scoville}, {Aussel}, {Brusa}, {Capak},
  {Carollo}, {Elvis}, {Giavalisco}, {Guzzo}, {Hasinger}, {Impey}, {Kneib},
  {LeFevre}, {Lilly}, {Mobasher}, {Renzini}, {Rich}, {Sanders}, {Schinnerer},
  {Schminovich}, {Shopbell}, {Taniguchi}, \& {Tyson}}]{2007ApJS..172....1S}
{Scoville}, N., {Aussel}, H., {Brusa}, M., {et~al.} 2007, \apjs, 172, 1

\bibitem[{{Shibuya} {et~al.}(2021){Shibuya}, {Miura}, {Iwadate}, {Fujimoto},
  {Harikane}, {Toba}, {Umayahara}, \& {Ito}}]{2021arXiv210603728S}
{Shibuya}, T., {Miura}, N., {Iwadate}, K., {et~al.} 2021, arXiv e-prints,
  arXiv:2106.03728

\bibitem[{{Speagle} {et~al.}(2014){Speagle}, {Steinhardt}, {Capak}, \&
  {Silverman}}]{2014ApJS..214...15S}
{Speagle}, J.~S., {Steinhardt}, C.~L., {Capak}, P.~L., \& {Silverman}, J.~D.
  2014, \apjs, 214, 15

\bibitem[{{Stark}(2016)}]{2016ARA&A..54..761S}
{Stark}, D.~P. 2016, \araa, 54, 761

\bibitem[{{Stefanon} {et~al.}(2017){Stefanon}, {Labb{\'e}}, {Bouwens},
  {Brammer}, {Oesch}, {Franx}, {Fynbo}, {Milvang-Jensen}, {Muzzin},
  {Illingworth}, {Le F{\`e}vre}, {Caputi}, {Holwerda}, {McCracken}, {Smit}, \&
  {Magee}}]{2017ApJ...851...43S}
{Stefanon}, M., {Labb{\'e}}, I., {Bouwens}, R.~J., {et~al.} 2017, \apj, 851, 43

\bibitem[{{Stefanon} {et~al.}(2019){Stefanon}, {Labb{\'e}}, {Bouwens}, {Oesch},
  {Ashby}, {Caputi}, {Franx}, {Fynbo}, {Illingworth}, {Le F{\`e}vre},
  {Marchesini}, {McCracken}, {Milvang-Jensen}, {Muzzin}, \& {van
  Dokkum}}]{2019ApJ...883...99S}
---. 2019, \apj, 883, 99

\bibitem[{{Steidel} {et~al.}(1999){Steidel}, {Adelberger}, {Giavalisco},
  {Dickinson}, \& {Pettini}}]{1999ApJ...519....1S}
{Steidel}, C.~C., {Adelberger}, K.~L., {Giavalisco}, M., {Dickinson}, M., \&
  {Pettini}, M. 1999, \apj, 519, 1

\bibitem[{{Stevans} {et~al.}(2018){Stevans}, {Finkelstein}, {Wold},
  {Kawinwanichakij}, {Papovich}, {Sherman}, {Ciardullo}, {Florez}, {Gronwall},
  {Jogee}, {Somerville}, \& {Yung}}]{2018ApJ...863...63S}
{Stevans}, M.~L., {Finkelstein}, S.~L., {Wold}, I., {et~al.} 2018, \apj, 863,
  63

\bibitem[{{Tacchella} {et~al.}(2018){Tacchella}, {Bose}, {Conroy},
  {Eisenstein}, \& {Johnson}}]{2018ApJ...868...92T}
{Tacchella}, S., {Bose}, S., {Conroy}, C., {Eisenstein}, D.~J., \& {Johnson},
  B.~D. 2018, \apj, 868, 92

\bibitem[{{Tacchella} {et~al.}(2021){Tacchella}, {Finkelstein}, {Bagley},
  {Dickinson}, {Ferguson}, {Giavalisco}, {Graziani}, {Grogin}, {Hathi},
  {Hutchison}, {Jung}, {Koekemoer}, {Larson}, {Papovich}, {Pirzkal},
  {Rojas-Ruiz}, {Song}, {Schneider}, {Somerville}, {Wilkins}, \&
  {Yung}}]{2021arXiv211105351T}
{Tacchella}, S., {Finkelstein}, S.~L., {Bagley}, M., {et~al.} 2021, arXiv
  e-prints, arXiv:2111.05351

\bibitem[{{Tamura} {et~al.}(2019){Tamura}, {Mawatari}, {Hashimoto}, {Inoue},
  {Zackrisson}, {Christensen}, {Binggeli}, {Matsuda}, {Matsuo}, {Takeuchi},
  {Asano}, {Sunaga}, {Shimizu}, {Okamoto}, {Yoshida}, {Lee}, {Shibuya},
  {Taniguchi}, {Umehata}, {Hatsukade}, {Kohno}, \& {Ota}}]{2019ApJ...874...27T}
{Tamura}, Y., {Mawatari}, K., {Hashimoto}, T., {et~al.} 2019, \apj, 874, 27

\bibitem[{{Tanaka} {et~al.}(2019){Tanaka}, {Valentino}, {Toft}, {Onodera},
  {Shimakawa}, {Ceverino}, {Faisst}, {Gallazzi}, {G{\'o}mez-Guijarro}, {Kubo},
  {Magdis}, {Steinhardt}, {Stockmann}, {Yabe}, \& {Zabl}}]{2019ApJ...885L..34T}
{Tanaka}, M., {Valentino}, F., {Toft}, S., {et~al.} 2019, \apjl, 885, L34

\bibitem[{{Treu} {et~al.}(2017){Treu}, {Abramson}, {Bradac}, {Brammer},
  {Fontana}, {Henry}, {Hoag}, {Huang}, {Mason}, {Morishita}, {Pentericci}, \&
  {Wang}}]{2017jwst.prop.1324T}
{Treu}, T.~L., {Abramson}, L.~E., {Bradac}, M., {et~al.} 2017, {Through the
  Looking GLASS: A JWST Exploration of Galaxy Formation and Evolution from
  Cosmic Dawn to Present Day}, JWST Proposal ID 1324. Cycle 0 Early Release
  Science

\bibitem[{{Valentino} {et~al.}(2020){Valentino}, {Tanaka}, {Davidzon}, {Toft},
  {G{\'o}mez-Guijarro}, {Stockmann}, {Onodera}, {Brammer}, {Ceverino},
  {Faisst}, {Gallazzi}, {Hayward}, {Ilbert}, {Kubo}, {Magdis}, {Selsing},
  {Shimakawa}, {Sparre}, {Steinhardt}, {Yabe}, \& {Zabl}}]{2020ApJ...889...93V}
{Valentino}, F., {Tanaka}, M., {Davidzon}, I., {et~al.} 2020, \apj, 889, 93

\bibitem[{{Walter} {et~al.}(2018){Walter}, {Riechers}, {Novak}, {Decarli},
  {Ferkinhoff}, {Venemans}, {Ba{\~n}ados}, {Bertoldi}, {Carilli}, {Fan},
  {Farina}, {Mazzucchelli}, {Neeleman}, {Rix}, {Strauss}, {Uzgil}, \&
  {Wang}}]{2018ApJ...869L..22W}
{Walter}, F., {Riechers}, D., {Novak}, M., {et~al.} 2018, \apjl, 869, L22

\bibitem[{{Wang} {et~al.}(2021){Wang}, {Yang}, {Fan}, {Hennawi}, {Barth},
  {Banados}, {Bian}, {Boutsia}, {Connor}, {Davies}, {Decarli}, {Eilers},
  {Farina}, {Green}, {Jiang}, {Li}, {Mazzucchelli}, {Nanni}, {Schindler},
  {Venemans}, {Walter}, {Wu}, \& {Yue}}]{2021ApJ...907L...1W}
{Wang}, F., {Yang}, J., {Fan}, X., {et~al.} 2021, \apjl, 907, L1

\bibitem[{{Wang} {et~al.}(2019){Wang}, {Schreiber}, {Elbaz}, {Yoshimura},
  {Kohno}, {Shu}, {Yamaguchi}, {Pannella}, {Franco}, {Huang}, {Lim}, \&
  {Wang}}]{2019Natur.572..211W}
{Wang}, T., {Schreiber}, C., {Elbaz}, D., {et~al.} 2019, \nat, 572, 211

\bibitem[{{Weaver} {et~al.}(2021){Weaver}, {Kauffmann}, {Ilbert}, {McCracken},
  {Moneti}, {Toft}, {Brammer}, {Shuntov}, {Davidzon}, {Hsieh}, {Laigle},
  {Anastasiou}, {Jespersen}, {Vinther}, {Capak}, {Casey}, {McPartland},
  {Milvang-Jensen}, {Mobasher}, {Sanders}, {Zalesky}, {Arnouts}, {Aussel},
  {Dunlop}, {Faisst}, {Franx}, {Furtak}, {Fynbo}, {Gould}, {Greve}, {Gwyn},
  {Kartaltepe}, {Kashino}, {Koekemoer}, {Kokorev}, {Le Fevre}, {Lilly},
  {Masters}, {Magdis}, {Mehta}, {Peng}, {Riechers}, {Salvato}, {Sawicki},
  {Scarlata}, {Scoville}, {Shirley}, {Sneppen}, {Smolcic}, {Steinhardt},
  {Stern}, {Tanaka}, {Taniguchi}, {Teplitz}, {Vaccari}, {Wang}, \&
  {Zamorani}}]{2021arXiv211013923W}
{Weaver}, J.~R., {Kauffmann}, O.~B., {Ilbert}, O., {et~al.} 2021, arXiv
  e-prints, arXiv:2110.13923

\bibitem[{{Williams} {et~al.}(2021){Williams}, {Oesch}, {Barrufet}, {Bezanson},
  {Bowler}, {Brammer}, {Dayal}, {Franx}, {Hutter}, {Labbe}, {Maseda}, {Ucci},
  \& {Whitaker}}]{2021jwst.prop.2514W}
{Williams}, C.~C., {Oesch}, P., {Barrufet}, L., {et~al.} 2021, {PANORAMIC - A
  Pure Parallel Wide Area Legacy Imaging Survey at 1-5 Micron}, JWST Proposal.
  Cycle 1

\bibitem[{{Willott} {et~al.}(2010){Willott}, {Albert}, {Arzoumanian},
  {Bergeron}, {Crampton}, {Delorme}, {Hutchings}, {Omont}, {Reyl{\'e}}, \&
  {Schade}}]{2010AJ....140..546W}
{Willott}, C.~J., {Albert}, L., {Arzoumanian}, D., {et~al.} 2010, \aj, 140, 546

\bibitem[{{Windhorst} {et~al.}(2017){Windhorst}, {Alpaslan}, {Ashcraft},
  {Broadhurst}, {Coe}, {Cohen}, {Conselice}, {Diego}, {Driver}, {Duncan},
  {Finkelstein}, {Frye}, {Grogin}, {Hathi}, {Hopkins}, {Jansen}, {Joshi},
  {Keel}, {Kelly}, {Kim}, {Koekemoer}, {Larson}, {Livermore}, {Marshall},
  {Mechtley}, {Pirzkal}, {Riess}, {Robotham}, {Rodney}, {Rottgering},
  {Rutkowski}, {Ryan}, {Smith}, {Straughn}, {Strolger}, {Tilvi}, {Wilkins},
  {Wyithe}, {Yan}, \& {Zitrin}}]{2017jwst.prop.1176W}
{Windhorst}, R.~A., {Alpaslan}, M., {Ashcraft}, T., {et~al.} 2017, {JWST
  Medium-Deep Fields - Windhorst IDS GTO Program}, JWST Proposal. Cycle 1

\bibitem[{{Yang} {et~al.}(2020){Yang}, {Wang}, {Fan}, {Hennawi}, {Davies},
  {Yue}, {Banados}, {Wu}, {Venemans}, {Barth}, {Bian}, {Boutsia}, {Decarli},
  {Farina}, {Green}, {Jiang}, {Li}, {Mazzucchelli}, \&
  {Walter}}]{2020ApJ...897L..14Y}
{Yang}, J., {Wang}, F., {Fan}, X., {et~al.} 2020, \apjl, 897, L14

\bibitem[{{Yung} {et~al.}(2019){Yung}, {Somerville}, {Finkelstein}, {Popping},
  \& {Dav{\'e}}}]{2019MNRAS.483.2983Y}
{Yung}, L.~Y.~A., {Somerville}, R.~S., {Finkelstein}, S.~L., {Popping}, G., \&
  {Dav{\'e}}, R. 2019, \mnras, 483, 2983

\bibitem[{{Yung} {et~al.}(2020){Yung}, {Somerville}, {Finkelstein}, {Popping},
  {Dav{\'e}}, {Venkatesan}, {Behroozi}, \& {Ferguson}}]{2020MNRAS.496.4574Y}
{Yung}, L.~Y.~A., {Somerville}, R.~S., {Finkelstein}, S.~L., {et~al.} 2020,
  \mnras, 496, 4574

\end{thebibliography}

\end{document}